\documentclass[useAMS,usenatbib]{mn2e}
\usepackage{times}
\usepackage[pdftex]{graphicx}
\usepackage{amsmath,amssymb}
\usepackage{newlfont,verbatim,enumerate,array}

\def\href#1#2{#2}
\def\urlinner#1{#1\endgroup}
\def\url{\begingroup\def\do##1{\catcode`##1 12 }%
\do\\\do\$\do\&\do\#\do\^\do\_\do\%\do\~ \ttfamily \urlinner}

\newcommand\vect[1]{\ensuremath{\mathbf{#1}}}
\newcommand\patspeed{km~s$^{-1}$~kpc$^{-1}$}
\newcommand\spiralampl{km$^2$~s$^{-2}$~kpc$^{-1}$}
\newcommand\amuse{\textit{AMUSE}}
\newcommand\bridge{\textit{Bridge}}
\newcommand\rotbridge{\textit{Rotating Bridge}}

\newcommand\kms{\ensuremath{\text{km~s}^{-1}}}
\newcommand\migration{\ensuremath{R_\mathrm{p}-R_\mathrm{b}}}
\newcommand\pmigration{\ensuremath{p(R_\mathrm{p}-R_\mathrm{b})}}
\newcommand\probio{\ensuremath{P_\mathrm{i-o}}}
\newcommand\proboi{\ensuremath{P_\mathrm{o-i}}}
\newcommand\olrbar{\ensuremath{\text{OLR}_\mathrm{bar}}}
\newcommand\crsp{\ensuremath{\text{CR}_\mathrm{sp}}}
\newcommand\olrsp{\ensuremath{\text{OLR}_\mathrm{sp}}}
\newcommand\ilrsp{\ensuremath{\text{ILR}_\mathrm{sp}}}

\title[Radial Migration of the Sun in The Milky Way]{Radial Migration of the Sun in the Milky
Way: a Statistical Study}
\author[C.A.\ Mart\'inez-Barbosa et al.]{C.A.\ Mart\'inez-Barbosa$^{1}$\thanks{E-mail:
cmartinez@strw.leidenuniv.nl} A.G.A.\ Brown$^{1}$\thanks{E-mail:
brown@strw.leidenuniv.nl} and S.\ Portegies Zwart $^{1}$\thanks{E-mail: spz@strw.leidenuniv.nl}\\
$^{1}$ Leiden Observatory, Leiden University, P.O.\ Box 9513 Leiden, 2300 RA, the Netherlands }
\begin{document}

\date{Accepted XXXXXXXXXXX. Received XXXXXXXXXXXX; in original form XXXXXXXXXXXXX}

\pagerange{\pageref{firstpage}--\pageref{lastpage}} \pubyear{2002}

\maketitle

\label{firstpage}

\begin{abstract}
The determination of the birth radius of the Sun is important to understand the evolution and consequent disruption of the Sun's birth cluster in the Galaxy. Motivated by this fact, we study the motion of the Sun in the Milky Way during the last $4.6$ Gyr in order to find its birth radius. We carried out orbit integrations backward in time using an analytical model of the Galaxy which includes the contribution of spiral arms and a central bar. We took into account the uncertainty in the parameters of the Milky Way potential as well as the uncertainty in the present day position and velocity of the Sun. We find that in general the Sun has not migrated from its birth place to its current position in the Galaxy $(R_\odot)$. However, significant radial migration of the Sun is possible when: 1) The $2:1$ Outer Lindblad resonance of the bar is separated from the corrotation resonance of spiral arms by a distance $\sim 1$~kpc. 2) When these two resonances are at the same Galactocentric position and further than the solar radius. In both cases the migration of the Sun is  from outer regions of the Galactic disk to $R_\odot$, placing the Sun's birth radius at around $11$ kpc. We find that in general it is unlikely that the Sun has migrated significantly from the inner regions of the Galactic disk to $R_\odot$.

\end{abstract}

\begin{keywords}
Galaxy: kinematics and dynamics -- open clusters and associations: general -- solar neighborhood --Sun:  general
\end{keywords}

\section{Introduction}\label{sec:intro} 

The study of the history of the Sun's motion within the Milky Way gravitational field is of great
interest to the understanding of the origins and evolution of the solar system \citep{adams} and the
study of past climate change and extinction of species on the earth \citep{fabo}. The determination
of the birth radius of the Sun is of particular interest in the context of radial migration and in
the quest for the siblings of the Sun \citep{brown, portegies09}. The work in this paper is
motivated by the possibility in the near future of combining large amounts of phase space data
collected by the {\em Gaia} mission \citep{gaia} with data on the chemical compositions of stars
(such as collected by the Gaia-ESO survey \citep{gaia-eso}) in order to search for the remnants of the Sun's
birth cluster. Our approach is to guide the search for the Sun's siblings by understanding in detail
the process of cluster disruption in the Galactic potential, using state of the art simulations. One
of the initial conditions of such simulations is the birth location, in practice the birth radius,
of the Sun's parent cluster. In this paper we present a parameter study of the Sun's past orbit in a
set of fully analytical Galactic potentials and we determine the most likely birth radius of the Sun
and by how much the Sun might have migrated radially within the Milky Way over its lifetime.

The displacement of stars from their birth radii is a process called radial migration. This can be
produced by different processes: interaction with transient spiral structure \citep{sellwood,
minchev06, roskar}, overlap of the dynamical resonances corresponding to the bar and spiral
structure \citep{minchev10, minchev11}, interference between spiral density waves that produce short
lived density peaks \citep{comparetta}, and interaction of the Milky Way disk with in-falling
satellites \citep{quillen09, bird}.

Since radial migration is a natural process in the evolution of Galactic disks, it is very likely
that the Sun has migrated from its formation place to its current position in the Galaxy.
\cite{wielen96} argued that the Sun was born at a Galactocentric distance of $6.6\pm 0.9$ kpc;
roughly 2 kpc nearer to the Galactic centre. He based his conclusions on the observation that the
Sun is more metal rich by $0.2$ dex with respect to most stars of the same age and Galactocentric
position \citep{holmberg} and the presence of a radial metallicity gradient in the Milky Way. Other
studies also support the idea that the Sun has migrated from its birth place. Based on
chemo-dynamical simulations of Galactic disks, \cite{minchev13} found that the most likely region in
which the Sun was born is between $4.4$ and $7.7$ kpc from the Galactic centre. 

However, if the metallicity of the Sun is not unusual with respect to the surrounding stars of the
same age it would no longer be valid to assume that the Sun migrated from the inner parts of our
Galaxy. By improving the accuracy in the determination of the effective temperature of the stars in
the data of the Geneva-Copenhagen Survey, \cite{casagrande} found that those stars are on average
100 K hotter and, hence, $0.1$ dex more metal rich. This result shifts the peak of the metallicity
distribution function to around the solar value, thus casting doubt on the observation that the Sun
is metal rich with respect to its surroundings. Further studies also support the idea that the Sun
is not an unusual star \citep{gustafsson98, gustafsson08, gustafsson10}.

The idea that the Sun might not have migrated considerably has been explored by several authors. By
solving the equations of motion of the Sun under the influence of a disk, a dark matter halo, spiral
arms, and the Galactic bar described by a multi-polar term, \cite{klacka} found that the radial
distance of the Sun varied between $7.6$ and $8.1$ kpc. They find migration only when the Sun
co-rotates with the spiral arms and when these structures represent very strong perturbations. On
the other hand, by using the method suggested by \cite{wielen96}, \cite{mishurov} found that the Sun
might have been born at approximately $7.4$ kpc from the Galactic centre.

Has the Sun migrated considerably? And if so, what are the conditions that allow such radial
migration? One way of solving these questions is by computing the motion of the Sun in the Galaxy
backwards in time. \cite{portegies09} used this technique to find that the Sun was born at a
distance of $r= 9.4$ kpc with respect to the Galactic centre. He used an axisymmetric potential for
modelling the Milky Way, which is not realistic and furthermore, he did not take into account the
uncertainty in the current position and velocity of the Sun (with respect to the Galactic reference
frame).

The aim of this paper is to address the question of the Sun's birth radius by carrying out orbit
integrations backward in time, using a more realistic model for the Galaxy which includes the
contribution of spiral arms and a central bar. We account for the uncertainty in the parameters of
the Milky Way potential as well as the uncertainty in the present day position and velocity the Sun.
The resulting parameter study is used to  obtain a statistical estimation of the Sun's
birth radius 4.6 Gyr ago. We use the \amuse\ framework \citep{portegies13} to perform our
computations.

This paper is organized as follows: in section \ref{sec:model} we describe the model that we use for
the Milky Way. In section \ref{sec:amuse} we provide a brief overview of the \amuse\ framework and
the modules we developed to compute potential past orbits of the Sun in the Galaxy. In section
\ref{sec:backtracing} we present the methodology to survey possible past orbits of the Sun and
thereby constrain its birth radius. In section \ref{sec:results} we analyse the orbit integration
results and address the question of whether or not the Sun has migrated in the Galaxy and the
conditions that would allow a considerable radial migration. In section \ref{sec:discussion} we
discuss the results and in section \ref{sec:summary} we present our conclusions and final remarks. 

\section{Galactic model}\label{sec:model}

Since the past history of the structure of the Milky Way is unknown, we simply assume that the
values of the Galactic parameters have been the same during the last $4.6$ Gyr, i.e.\ during the
lifetime of the Sun \citep{buonano}. We model the Milky Way as a fully analytical potential
 that contains an
axisymmetric component together with a rotating central bar and spiral arms. We use the potentials
and parameters of \cite{allen} to model the axisymmetric part of the Galaxy, which consist of a
central bulge, a disk and a dark matter halo. The values of the parameters of these Galactic
components are shown in table \ref{tab:galparams}. For the central bar and spiral arms we use the models
presented in \cite{merce2} and \cite{antoja} as detailed below.

\begin{table}
  \caption{Parameters of the Milky Way model potential.}
  \label{tab:galparams}
  \begin{tabular}{ll} \hline
    \multicolumn{2}{c}{\textbf{ Axisymmetric component}} \\ \hline 
    Mass of the bulge ($M_\mathrm{b}$) & $1.41\times 10^{10}$ $M_{\odot}$ \\ 
    Scale length bulge ($b_\mathrm{1}$) & $0.3873$ kpc\\
    Disk mass ($M_\mathrm{d}$) & $8.56\times10^{10}$ $M_{\odot}$\\
    Scale length disk 1 ($a_\mathrm{2}$) & $5.31$ kpc\\
    Scale length disk 2 ($b_\mathrm{2}$) & $0.25$ kpc\\
    Halo mass ($M_\mathrm{h}$) & $1.07\times 10^{11} $ $M_{\odot}$\\
    Scale length halo ($a_\mathrm{3}$) & 12 kpc\\ \hline 
    \multicolumn{2}{c}{\textbf{Central Bar}} \\ \hline 
    Pattern speed ($\Omega_\mathrm{bar}$) & $40$--$70$ km~s$^{-1}$~kpc$^{-1}$\\ 
    Semi-major axis ($a$) & $3.12$ kpc \\
    Axis ratio ($b/a$) & $0.37$ \\
    Mass ($M_\mathrm{bar}$) & $9.8\times10^9$--$1.4\times10^{10}$ $M_{\odot}$\\ 
    Orientation & $20^\circ$ \\ \hline
    \multicolumn{2}{c}{\textbf{ Spiral arms}} \\ \hline 
    Pattern speed ($\Omega_\mathrm{sp}$) & $15$--$30$ km~s$^{-1}$~kpc$^{-1}$\\
    Locus beginning ($R_\mathrm{{sp}}$) & $3.12$ kpc\\
    Number of spiral arms ($m$) & $2$, $4$\\
    Spiral amplitude ($A_\mathrm{sp}$) &  $650$--$1300$ \spiralampl  \\
    Strength of the spiral arms ($\epsilon$) & $0.02$-- $0.06$ \\
    Pitch angle ($i$) & $ 12.8^\circ$ \\
    Scale length ($R_\mathrm{{\Sigma}}$) & $2.5$ kpc\\
    Orientation & $20 ^\circ$ \\
    		        \hline
  \end{tabular}
\end{table}

\subsection{Central bar}\label{sec:cbar}

The central bar of the Milky Way is modelled as a Ferrers bar \citep{ferrers} which is described by
a density distribution of the form:

\begin{equation}
  \rho_\mathrm{bar}= 
  \begin{cases}
    \rho_0 \left( 1-n^2 \right)^k & n < 1\\
    0 & n \geq 1
  \end{cases}\,,
\end{equation}
where $n^2= x^2/a^2 + y^2/b^2$ determines the shape of the bar potential, where $a$ and $b$ are the
semi-major and semi-minor axes of the bar, respectively. Here, $x$ and $y$ are the axes of a frame that corrotates with the bar. $\rho_0$ represents the central density
of the bar and the parameter $k$ measures the degree of concentration of the bar. Larger values of $k$
correspond to a more concentrated the bar. The extreme case of a constant density bar is obtained
for $k=0$ \citep{merce1}. Following \cite{merce2} we use $k=1$. For these models the mass of the bar
is given by:
\begin{equation}
M_\mathrm{bar}= \frac{2^{(2k+3)}\pi ab^2\rho_0\Gamma(k+1)\Gamma(k+2) }{\Gamma(2k+4)}\, ,
\end{equation}
where $\Gamma$ is the Gamma function.

\subsubsection{Galactic bar parameters}

\paragraph*{Number of bars} The inner part of the Galaxy has been extensively studied within the
\textit{COBE/DIRBE} \citep{weiland} and \textit{Spitzer/GLIMPSE} \citep{churchwell} projects, which
demonstrated that the centre of the Milky Way is a complex structure. While the \textit{COBE/DIRBE}
data showed that the surface brightness distribution of the bulge resembles a flattened ellipse with
a minor-to-major axis ratio of $\sim0.6$, the \textit{Spitzer/GLIMPSE} survey confirmed the
existence of a second bar \citep{benjamin} which was previously observed by \cite{hammersley}. Since
the longitude and length ratios of these bars are in strong disagreement with both simulations and
observations, \cite{merce2} suggested that there is only a single bar at the centre of the Milky
Way, which was confirmed by the analysis of \cite{martinez}, who show that the observations of the
central region of the Milky Way can be explained by one bar. Hence we take into account the
contribution of only one bar in the potential model of the Milky Way, using the parameters as
obtained from the \textit{COBE/DIRBE} survey. 

\paragraph*{Pattern speed} The value of the pattern speed of the bar is uncertain.  From theoretical and observational data \cite{dehnen} concluded that  $\Omega_\mathrm{bar}= 50\pm 3$~ \patspeed; however, \cite{bissantz}  argued that a more suitable value for the pattern speed of the bar is $60\pm5$ \patspeed\ . Taking into account these values, we assume that the bar rotates as a
rigid body with a pattern speed between 40 and 70 \patspeed.

\paragraph*{Semi-major axis and axis ratio} Based on the best fit model by \cite{freuderich} and on
the uncertainty in the current solar Galactocentric position\footnote{We conservatively assume the
uncertainty in the distance from the Sun to the Galactic centre is $0.5$ kpc}, the semi-major axis
of the \textit{COBE/DIRBE} bar is between $2.96$ and $3.31$ kpc. With these assumptions the axis
ratio of the bar is between $0.36$ and $0.38$. In our simulations we maintain these two parameters
constant with the values listed in table \ref{tab:galparams}. 

\paragraph*{Mass and orientation of the bar} Several studies suggest that the mass of the
\textit{COBE/DIRBE} bar is in the range $0.98$--$2$$\times 10^{10}$ $M_{\odot}$
\citep{weiner, dwek, matsumoto, zhao}.  Given that the bar is formed from the bulge, we  assume the mass of the bar is in the range $9.8\times 10^9-1.4\times 10^{10}$~$M_{\odot}$.

The orientation of the bar is defined as the angle between its major axis and the line that
joins the Galactic centre with the current position of the Sun. We fixed this angle at $20^\circ$
\citep{pichardo04, pichardo1, merce2}, as illustrated in Fig.\ \ref{fig:config}.

\paragraph*{Effect of a growing bar} From $N$-body simulations it appears that bars in galaxies are
formed during the first $1.4$ Gyr of their evolution \citep {fux,polyachenko}. Thus, we assume that
the bar was already present in the Milky Way when the Sun was formed $4.6$ Gyr ago.

\subsection{Spiral arms}\label{sec:sarms}

The spiral arms in our Milky Way Models are represented as periodic perturbations of the
axisymmetric potential. Following \cite{contopolus} the potential of such perturbations in the plane
is given by:
\begin{equation}
\phi_\mathrm{sp}= -A_\mathrm{sp}Re^{-R/R_{\Sigma}}\cos{\left(m(\phi)-g(R) \right)} \label{speq}\, ,
\end{equation} 
where $A_\mathrm{sp}$ is the amplitude of the spiral arms. $R$ and $\phi$ are the 
cylindrical coordinates of a star measured in a corotating frame with the spiral arms. $R_{\Sigma}$ and $m$ are the scale length and the number of spiral arms,
respectively. The function $g(R)$ defines the locus shape of the spiral arms. We use the same
prescription as \cite{antoja}:

\begin{equation}
g(R)= \left( \frac{m}{N\tan{i}} \right) \ln\left( 1+ \left( \frac{R}{R_\mathrm{sp}} \right)^N
  \right)\, .
\end{equation}

$N$ is a parameter which measures how sharply the change from a bar to a spiral structure occurs in
the inner regions of the Milky Way. $N \rightarrow \infty $ produces spiral arms that begin forming
an angle of $\sim 90^o$ with respect to the line that joins the two starting points of the locus
\citep{antoja} (as illustrated in Fig.\ \ref{fig:config} below). To approximate this case we use
$N=100$. $R_\mathrm{sp}$ is the separation distance of the beginning of the spiral shape locus and
$\tan{i}$ is the tangent of the pitch angle. 

\subsubsection{Spiral arm parameters}

\paragraph*{Pattern speed} Some studies point out that the spiral arms of the Milky Way
approximately rotate with a pattern speed $\Omega_\mathrm{sp}= 25 \pm 1$ \patspeed\
\citep[e.g.][]{dias}, while others argue that the value is $\Omega_\mathrm{sp}= 20$ \patspeed\
\citep[e.g.][]{martos}. Since the pattern speed of the spiral arms is uncertain, we chose a range
between 15 and 30 \patspeed, as in \cite{antoja}. In addition we assume the spiral arms rotate as
rigid bodies.

\paragraph*{Locus shape, starting point, and orientation of the spiral arms}  In the simulations we adopt the spiral arm model obtained from a fit to the Scutum and Perseus arms. This is the so-called `locus 2' in the work of \cite{antoja}.  We also assume that the spiral structure starts at the edges of the bar. Hence  $R_\mathrm{sp}= 3.12$~kpc. With this configuration the angle between the line connecting the starting point of the spiral arms and the Galactic centre-Sun line is  $20^\circ$ (see Fig.\ \ref{fig:config}).

\paragraph*{Number of spiral arms} \cite{drimmel00} used K-band photometry of the Galactic plane to
conclude that the Milky Way contains two spiral arms. On the other hand, \cite{valee02} reviewed a
number of studies about the spiral structure of the Galaxy --- mostly based on young stars, gas and
dust --- and he concluded that the best overall fit is provided by a four-armed spiral pattern.
Given this discrepancy, we carry out simulations with $m=2$ or $m=4$ spiral arms.

\paragraph*{Amplitude and strength of the spiral arms}  We used the amplitude of the spiral arms from the Locus 2 model in \cite{antoja}, which is between  650 and 1100~\spiralampl. The strength of the
spiral arms \citep[as defined in Sect.\ 5 of][]{antoja} corresponding to this range of amplitudes is
between $0.029$ and $0.05$.   We however explored the motion of the Sun for amplitudes of up to 1300~\spiralampl\ ($\epsilon \sim 0.06$) in a two-armed spiral structure.

\paragraph*{Other parameters} We also use the value of the locus 2 model of \cite{antoja} for the
pitch angle ($i$) and scale length ($R_\Sigma$) of the spiral perturbation. These values are listed
in table \ref{tab:galparams}.

\paragraph*{Transient spiral structure} Several theoretical studies support the idea that spiral
arms in galaxies are transient structures \citep{sellwood, sellwood11}. Nevertheless, \cite{fujii1}
found that spiral arms in pure stellar disks can survive for more than 10 Gyr when a sufficiently
large number of particles ($\sim 10^7$) is used in the simulations. In this work we use only static
spiral structure. 

\paragraph*{Multiple spiral patterns}  \cite{lepine10} have argued that the corrotation radius of the spiral arms  is located at solar radius, i.e. at $R=8.4$~kpc; however based on the orbits of the Hyades and coma Berenices  moving groups, \cite{quillen05} concluded that the 4:1 inner Lindblad resonance of the spiral arms  is located at  the solar position, placing the corrotation resonance at around 12 kpc. To reconcile the uncertainty in the location of the coronation resonance of the spiral structure, \cite{lepine11} suggested  the existence  of multiple spiral arms with different pattern speeds in the Galaxy. while the main grand-design spiral pattern has its corrotation at $8.4$~kpc, an outer $m=2$ pattern would have its corrotation resonance at about 12 kpc, with the 4:1 inner Lindblad resonance at the position of the Sun.  These multiple spiral patterns have been observed in N-body simulations \citep[See e.g.][]{quillen11}.  

In this work we also consider a superposition of spiral patterns as suggested by \cite{lepine11} to study the motion of the Sun in the Galaxy.

%_____________________________________________

\section{The {\amuse} framework}\label{sec:amuse}

{\amuse}, the Astrophysical MUltipurpose Software Environment \citep{portegies13}, is a framework
implemented in \textit{Python} in which different astrophysical simulation codes can be coupled to
evolve complex systems involving different physical processes. For example, one can couple an
$N$-body code with a stellar evolution code to create an open cluster simulation in which both
gravitational interactions and the evolution of the stars are included. Currently {\amuse} provides
interfaces to codes for gravitational dynamics, stellar evolution, hydrodynamics and radiative
transfer. 

{\amuse} is used by writing \textit{Python} scripts to access all the numerical codes and their
capabilities. Every code incorporated in {\amuse} can be used through a standard interface which is
defined depending on the domain of the code. For instance, a gravitational dynamics interface
defines how a system of particles moves with time and in this case, the user can add or remove
particles and update their properties. We created an interface in {\amuse} for the Galactic model
described in Sect.\ \ref{sec:model}. For details about how to use {\amuse} we
refer the reader to \cite{portegies13} and \cite{pelupessy13}. More information can be also found
at \texttt{http://amusecode.org}.

The computation of the stellar motion due to an external gravitational field can be done in {\amuse} through the {\bridge} \citep{bridge}
interface. This code uses a second-order Leapfrog method to compute the velocity of the stars due to
the gravitational field of the Galaxy. All these computations are performed in an inertial frame.
Given that the potentials of the bar and spiral arms are defined to be time independent in a
reference system that co-rotates either with the bar or with the spiral arms, we modified {\bridge}
to compute the position and velocity of the Sun in one of such non-inertial frames. Moreover, since
the time symmetry of the second-order Leapfrog is no longer valid in a rotating frame we need to
use a higher order scheme. These modifications resulted in a new interface called {\rotbridge}.  This code can also be used to perform self-consistent $N$-body simulations of stellar clusters  that also respond to the gravitational non-static force from their parent galaxies. In these simulations the internal cluster effects like self gravity and stellar evolution
can be taken into account. In Appendix \ref{app:bridge} we derive the equations of motion for the
{\rotbridge} for a single particle and its generalization to a system of self-interacting
particles. We also show the accuracy of this code under different Galactic parameters.

%____________________________________________________________
\section{Back-tracing the Sun's orbit}\label{sec:backtracing}

\begin{figure}
  \centering
  \includegraphics[width= 9cm, height= 8cm]{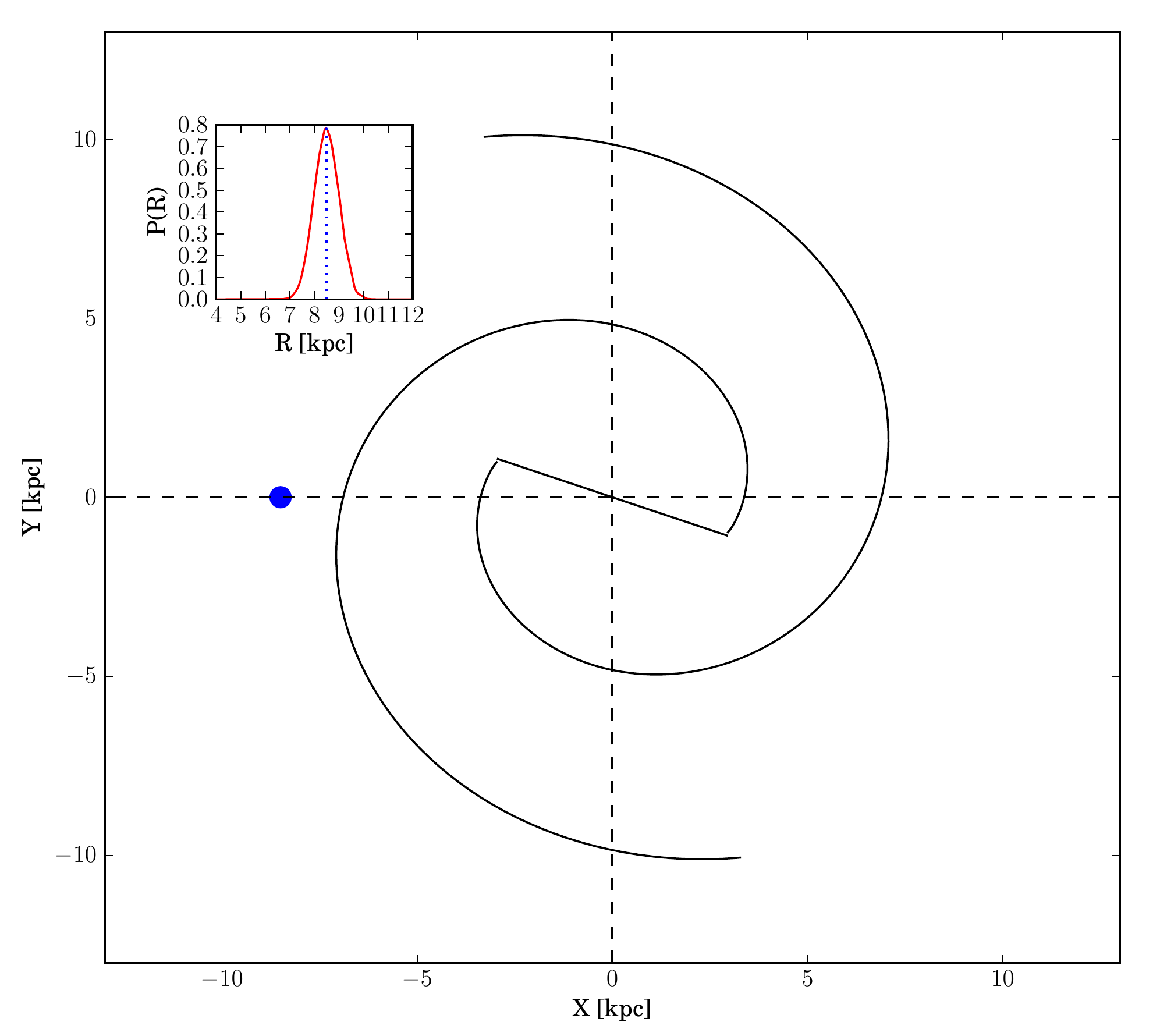}
  \caption{Configuration of the Galactic potential at the beginning of the backwards integration in
    time. The spiral arms are assumed to start at the ends of the major axis of the bar. The blue
    circle is the current position of the Sun,  $r_\odot= (-8.5, 0)$ kpc. The angle
    the Sun-Galactic centre line makes with respect to the semi-major axis of the bar is $20^\circ$.
    The inset shows the distribution of 5000 Galactocentric distances that were selected from a 3D
    Gaussian centred at the current phase-space coordinates of the Sun.\label{fig:config}}
\end{figure}

Contrary to the epicyclic trajectories that stars follow when they move under the action of an
axisymmetric potential, the orbits of stars become more complicated when the gravitational
fluctuations generated by the central bar and spiral arms are taken into account, specially where chaos might be important. In chaotic regions, small deviations in the initial position and/or velocity of stars produce significant variations in their final location. Hence, in order to determine the birth place of one star, it is necessary to use a precise numerical code able to resolve the substantial and sudden changes in acceleration that such star experiments. Additionally, it is necessary to compute its orbit  backwards in time by using a sampling of positions and velocities around the star's current (uncertain) location in phase-space.  With this last procedure we get statistical
information about the region in the Galaxy where the star might have been born. We follow this
methodology to find the most probable birth radius and velocity of the Sun to infer whether or
not it has radially migrated during its lifetime.  To ensure numerical accuracy in the orbit integration we used a 6th order Leapfrog in the {\rotbridge} with a time
step of $0.5$ Myr. This choice leads to a fractional energy error of the order of $10^{-10}$. (See Sect.\ \ref{sect:high_order}). 

As a first step we generate 5000 random positions and velocities which are within the measurement
uncertainties from the current Galactocentric position and velocity of the Sun $(\vect{r}_\odot,
\vect{v}_\odot)$. This selection was made from a 4D normal distribution centred at $(\vect{r}_\odot,
\vect{v}_\odot)$ with standard deviations ($\sigma$) corresponding to the measured errors in these
coordinates. We assume that the Sun is currently located at: $\vect{r}_\odot = (-R_\odot, 0)$ kpc;
where the distance of the Sun to the Galactic centre is $R_\odot \pm \sigma_R= 8.5\pm 0.5$~kpc. The uncertainty in $y_\odot$ is set to zero as the Sun is by definition located on the
$x$-axis of the Galactic reference frame.

Since we consider the motion of the Sun only on the Galactic plane, the velocity of the Sun is: $\vect{v}_\odot = (U_\odot, V_\odot)$, where:
\begin{align}
 U_\odot \pm \sigma_{U}  &= 11.1 \pm 1.2 \text{\ } \kms  \nonumber \\
 V_\odot \pm \sigma_{V} &= (12.4+ V_\mathrm{LSR}) \pm 2.1 \text{\ }\kms\,. 
\end{align}
The vector $(11.1\pm1.2, 12.4\pm 2.1)$ \kms\ is the peculiar motion of the Sun
\citep{schonrich} and $V_\mathrm{LSR}$ is the velocity of the local standard of rest which depends
on the Galactic parameters that are listed in table \ref{tab:galparams}. We use the conventional
Galactocentric Cartesian coordinate system. This means that translated to a Sun-centred reference
frame the $x$-axis points toward the Galactic centre, the $y$-axis in the direction of Galactic
rotation, and the $z$-axis completes the right-handed coordinate system.

Recently \cite{bovy12} found an offset between the  rotational velocity of the Sun and $V_\mathrm{LSR}$  of 
$26\pm 3$ km~s$^{-1}$, which is larger than the value measured by \cite{schonrich}. We also use this value to trace back the Sun's orbit.

In Fig.\ \ref{fig:config} we show the configuration of the Galactic potential at the beginning of
the backwards integration in time . Since it is unknown how spiral arms are oriented with respect to
the bar at the centre of the Galaxy, we assume that they start at the edges of the bar. The
blue circle in this Figure represents the current location of the Sun. The line
from the Sun to the Galactic centre makes an angle of $20^\circ$ with the semi-major axis of the
bar. In the small plot located at the left top of Fig.\ \ref{fig:config} we show the distribution of
the 5000 positions in cylindrical radius $R$.

Each of the 5000 positions and velocities that were generated from the 4D normal distribution are
used to construct a set of present-day phase space vectors with (cylindrical) coordinates:
$(R_\mathrm{p}, \varphi_\mathrm{p}, v_{R_\mathrm{p}},v_{\varphi_\mathrm{p}})_k$; $k= 1,\dots,5000$ (Note that $\varphi_\mathrm{p}$ is fixed at $\pi$). The Sun
is then located at each of these vectors and its orbit is computed backwards in time until $4.6$ Gyr
have elapsed. Before starting the integration we reversed the velocity components of the Sun as
well as the direction of rotation of the bar and spiral arms\footnote{The convention used in the
 {\rotbridge} is right-handed; hence, for the backward integration in time the pattern speed of the
bar and spiral arms are positive.}. 

After integrating the orbit of the Sun backwards in time we obtain a sample of birth phase-space
coordinates $(R_{\mathrm{b}}, \varphi_{\mathrm{b}}, v_{R_\mathrm{b}},
v_{\varphi_\mathrm{b}})_k$; $k=1,\dots,5000$. The distributions of present day and
birth phase space coordinates then allow us to study the past motion of the Sun and infer whether or
not it has migrated during its lifetime.

To take the uncertainties on the Galactic model into account we also varied the bar and spiral arm
parameters according to the values listed in table \ref{tab:galparams}. For a subset of the
Galactic model parameters we verified that 5000 birth phase-space coordinates are a representative
number for sampling the position and velocity of the Sun $4.6$ Gyr ago. By means of the
Kolmogorov-Smirnoff test, we found that the distribution of positions and velocities of the Sun
after integrating its orbit backwards in time, is the same when $k= 5000$, $10\,000$ or $20\,000$.
Depending on the Galactic parameters, the $p$-value from the test is between $0.2$ and $0.98$.

%______________________________

\section{Results} \label{sec:results}

\begin{figure}
  \centering
  \includegraphics[width= 5cm, height= 13.5 cm]{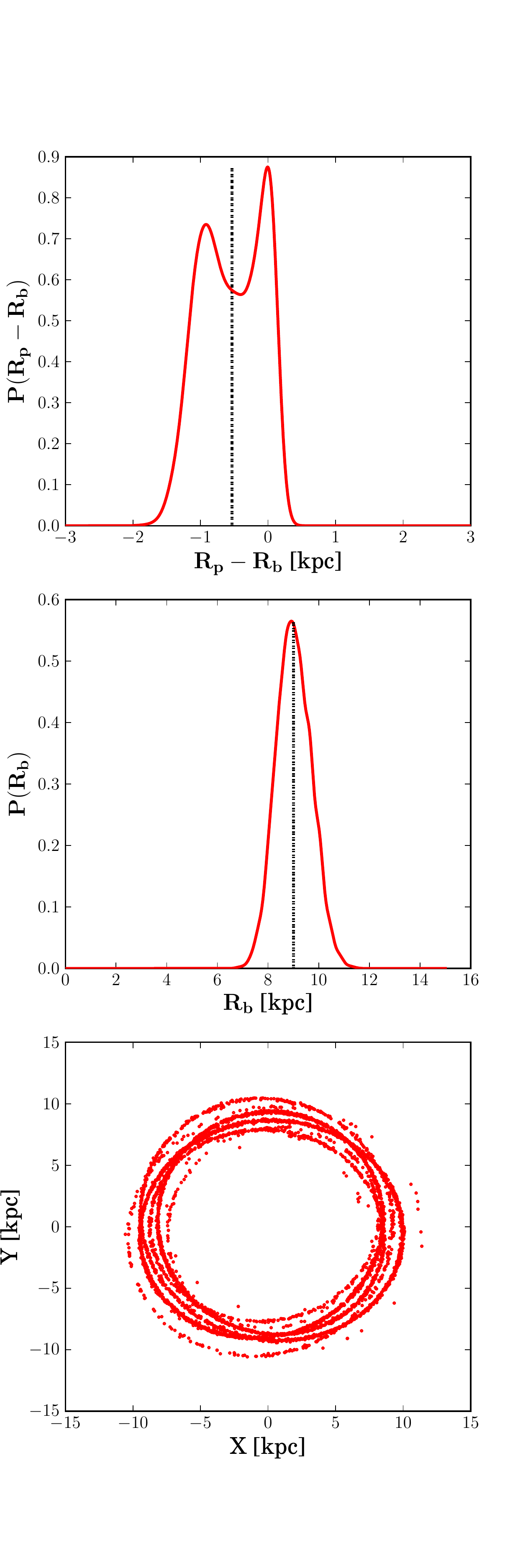}
  \caption{Results of the back-tracing of the Sun's orbit in a purely axisymmetric Milky Way
    potential. \textbf{Top:} the migration distribution \pmigration. \textbf{Middle:} distribution
    of the birth radius of the Sun $p(R_\mathrm{b})$. \textbf{Bottom:} the distribution of birth
    locations of the Sun on the $xy$-plane. The dotted black line in the top two panels
    represents the median of distributions. Note that this is negative for \pmigration, which means
    that the migration of the Sun is from outer regions of the galaxy to $R_\odot$. The distribution
    of birth positions of the Sun seen on the $xy$ plane suggest that it is not possible to
    determine the exact formation place of the Sun $4.6$ Gyr ago.\label{fig:axi}}
\end{figure}

 \begin{figure*}
  \centering
  \includegraphics[width= 18cm, height= 15cm]{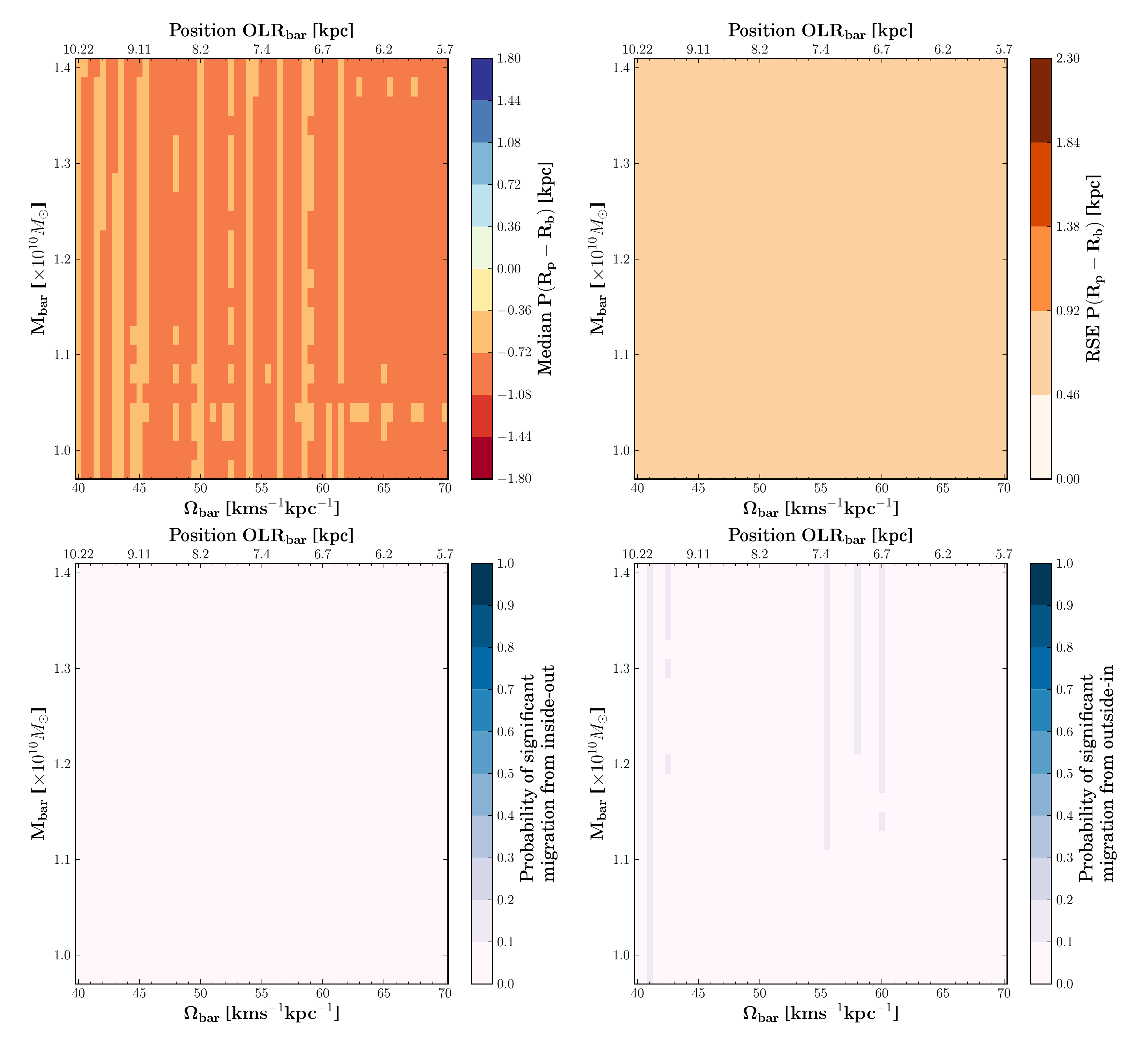}
  \caption{\textbf{Top:} Median and RSE of the migration distribution \pmigration\ as a function of
  the mass and pattern speed of the bar. Negative values in the median indicate migration from outer
  regions of the Galactic disk to $R_\odot$, while positive values indicate migration from inner
  parts to $R_\odot$. The position of the bar's outer Lindblad resonance, \olrbar, with
  respect to the Galactic centre is also shown. For this set of simulations the position of
  \crsp\ is fixed at $10.9$ kpc. \textbf{Bottom:} \probio\ and \proboi\ as a function of the
  mass and pattern speed of the bar.\label{fig:ri_bar_effect}}
\end{figure*}

For every choice of bar and spiral arm parameters we have the distribution of the present day phase
space coordinates of the Sun $p(\vect{r}_\mathrm{p},\vect{v}_\mathrm{p})$ and of the Sun's phase
space coordinates at birth $p(\vect{r}_\mathrm{b},\vect{v}_\mathrm{b})$. The amount of radial
migration experienced by the Sun during its motion through the Galaxy can be obtained from the
probability distribution \pmigration\ (referred to below as the `migration distribution') of the
difference in the radial distance between the present day and birth locations of the Sun. We use the
median of the distribution to decide whether or not the Sun has migrated a considerable  distance
during its lifetime:

\begin{enumerate}
  \item Median $\pmigration>d_\mathrm{m}$: the Sun migrated from inner regions of the Galactic
    disk to $R_\odot$ (migration from inside-out).
  \item Median $-d_\mathrm{m}\leq\pmigration\leq d_\mathrm{m}$: the Sun has not migrated
  \item Median $\pmigration<-d_\mathrm{m}$: the Sun migrated from outer regions of the Galactic
    disk to $R_\odot$ (migration from outside-in). 
\end{enumerate}

The parameter $d_\mathrm{m}$ indicates when the value of \migration\ is considered to indicate a
significant migration of the Sun within the Galaxy. We derive the value of $d_\mathrm{m}$ by
considering the distribution \pmigration\ for the case of a purely axisymmetric Galaxy, in which
case for the Sun's orbital parameters the migration should be limited. The migration distribution
for this case is shown in Fig.\ \ref{fig:axi}. From this distribution it can be seen that for the
axisymmetric case indeed the Sun migrates only little on average ($\sim0.6$ kpc) and that the
maximum migration distance is about $1.7$ kpc (note that $\pmigration=0$ for
$\migration\lesssim-1.7$ kpc). Based on this result we use $d_\mathrm{m}=1.7$ kpc in the discussions
of the results below. Considering changes in the Sun's radial distance larger than $1.7$~kpc
as significant migration is consistent with the estimates of the Sun's migration made by
\cite{wielen96} and \cite{minchev13}.

The value of the median of \pmigration\ is not enough to characterize this probability distribution
which is often multi-modal (see top panel of Fig.\ \ref{fig:axi}) and we thus introduce the
following quantities:
\begin{equation}
  \begin{aligned} 
   P_\mathrm{i-o} &= \int_{d_\mathrm{m}}^\infty \pmigration\,d(\migration) \\
   P_\mathrm{o-i} &= \int_{-\infty}^{-d_\mathrm{m}} \pmigration\,d(\migration)
  \end{aligned}\,,
  \label{eq:migration}
\end{equation}
where {\probio} is the probability that the Sun has experienced considerable migration from the
inner regions of the Galactic disk to its present day position, while \proboi\ is the
probability that the Sun has significantly migrated in the other direction. One of the aims of our
study is to find Milky Way potentials for which the above probabilities are substantial, thus
indicating that the Sun has likely migrated a considerable distance over its lifetime.

We also characterize the width of the distribution \pmigration\ through the so-called
Robust Scatter Estimate (RSE) \citep{lindegren} which is defined as $\mathrm{RSE} = 0.390152\times (P90-P10)$, where
$P10$ and $P90$ are the 10th and 90th percentiles of the distribution, and the numerical constant is
chosen to make the RSE equal to the standard deviation for a Gaussian distribution
  
The orbit integrations were carried out  by using the peculiar velocity of the Sun inferred by \cite{schonrich}, unless otherwise stated.

\subsection{Radial migration of the Sun as a function of bar parameters}\label{sec:bar}

\begin{figure}
  \centering
  \includegraphics[width= 5cm, height= 14cm]{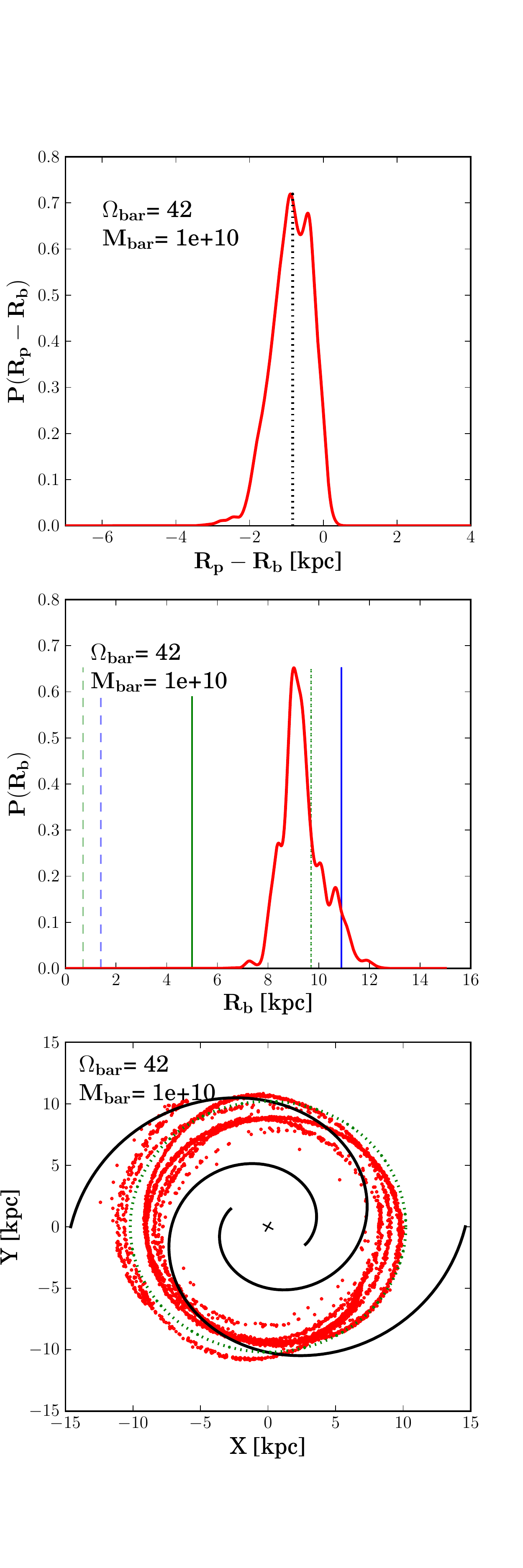}
  \caption{\textbf{Top:} Distribution function \pmigration\ for the Galaxy model with weak spiral
  arms and a central bar. The vertical dotted black line is the median of the distribution.
  \textbf{Middle:} Radial distribution of the birth radius of the Sun $p(R_\mathrm{b})$ for the
  same Galactic parameters. The vertical green lines represent the location of the resonances
  produced by the bar while the blue lines, represent the location of the resonances due to the
  spiral arms. The dashed, solid and dotted lines represent the 2:1 inner Lindblad (ILR),
  co-rotation (CR) and 2:1 outer Lindblad (OLR) resonances respectively. Hereafter, we will use this
  same convention. \textbf{Bottom:} Distribution of birth positions of the Sun seen on the $xy$
  plane. The \olrbar\ is shown as the circular dotted green line. We also show the
  configuration of the spiral arm potential $4.6$ Gyr ago.\label{fig:xy_bar_effect}}
\end{figure}

In order to study the radial migration of the Sun under the variation of mass and pattern speed of
the bar, we fixed the amplitude, pattern speed and number of spiral arms such that they have little
effect on the Sun's orbit. We chose the values: $A= 650$ \spiralampl, $\Omega_\mathrm{sp}=20$
\patspeed, and $m=2$.  With these values of amplitude and pattern speed we produce spiral arms with
a strength at the lowest limit ($\epsilon= 0.029$)  and resonances located in extreme regions of the
Galactic disk. The 2:1 inner/outer Lindblad resonance of the spiral arms (\ilrsp, \olrsp) and the
co-rotation resonance (\crsp), are located at $1.4$ kpc , $16$ kpc and $10.9$ kpc respectively.  

In Fig.\ \ref{fig:ri_bar_effect} we show the median, RSE, \probio, and \proboi\ of the distribution
\pmigration\ as a function of the mass and pattern speed of the bar. The mass of the bar was varied
in steps of $0.02$ $M_\odot$ and the pattern speed in steps of $0.5$ \patspeed. The maximum and
minimum values of $M_\mathrm{bar}$ and $\Omega_\mathrm{bar}$ were set according to the ranges listed
in table \ref{tab:galparams}. Fig.\ \ref{fig:ri_bar_effect} also shows the position of the 2:1 outer
Lindblad resonance of the bar (\olrbar). 

Note that the median of the distribution {\pmigration} is always negative. This indicates that the
migration of the Sun in this case on average is from outer regions of the Galactic disk to
$R_\odot$.  The median of \pmigration\ is also always lower than $1.08$~kpc, independently of the mass and pattern speed of the bar.

On the other hand from the bottom  panel of Fig.\ \ref{fig:ri_bar_effect} it is clear that
regardless of the mass and pattern speed of the bar, it is unlikely that the Sun has migrated
considerably from the inner or outer regions of the Galactic disk to $R_\odot$. The low probability of
significant radial migration can also be seen in the width of the migration distribution which is always
below $0.92$ kpc (top right panel Fig.\ \ref{fig:ri_bar_effect}).

We conclude that the presence of the central bar of the Milky Way does not produce
considerable radial migration of the Sun. This result is not surprising, because although the
\olrbar\ has played an important role in shaping the stellar velocity distribution function in the
solar neighbourhood \citep{dehnen, minchev}, the gravitational force produced by the bar falls
steeply with radius, reaching about 1\% of its total value at $R_\odot$ \citep{dehnen}.
\cite{klacka} studied the motion of the Sun in an analytical model of the Galaxy that considers a
multipolar expansion of the bar potential. By assuming the current location of the Sun  as
$\vect{r}_{\odot}= (-8, 0, 0)$ kpc and $\vect{v}_\odot= (0, 220, 0)$ \kms,  they
found that the central bar of the Galaxy does not generate considerable radial migration of the Sun
if spiral arms are not considered, changing the Galactocentric distance of the Sun only 1\% from its
current value $R_\odot$. We find more than 1\% change in radius because we take into account the
potential of the spiral arms in the Galactic model.

Figure \ref{fig:xy_bar_effect} shows the distributions \pmigration\ and $p(R_\mathrm{b})$ for a
choice of bar parameters. In this
specific case the median of \pmigration\ is $-0.83$ kpc, which means that the birth radius of the
Sun is around $9.3$~kpc. From the distribution of Sun's possible birth positions  on the $xy$
plane (bottom panel Fig.\ \ref{fig:xy_bar_effect}) it is clear that even for this smooth and static
potential only the birth radius of the Sun can be constrained. The uncertainty in $\varphi$ for the
Sun's birth location is caused by the uncertainty in the present day phase space coordinates of the
Sun. 

In this Section we have simulated the  radial migration of the Sun as a function of mass and pattern
speed of the bar. We find no significant migration. In the next Section we study the motion of the Sun when the parameters of the spiral arms  are varied.

\subsection{Radial migration of the Sun as a function of spiral arm
parameters}\label{sec:spiral_effects}

\begin{figure*}
  \centering
  \includegraphics[width= 18 cm, height= 15 cm]{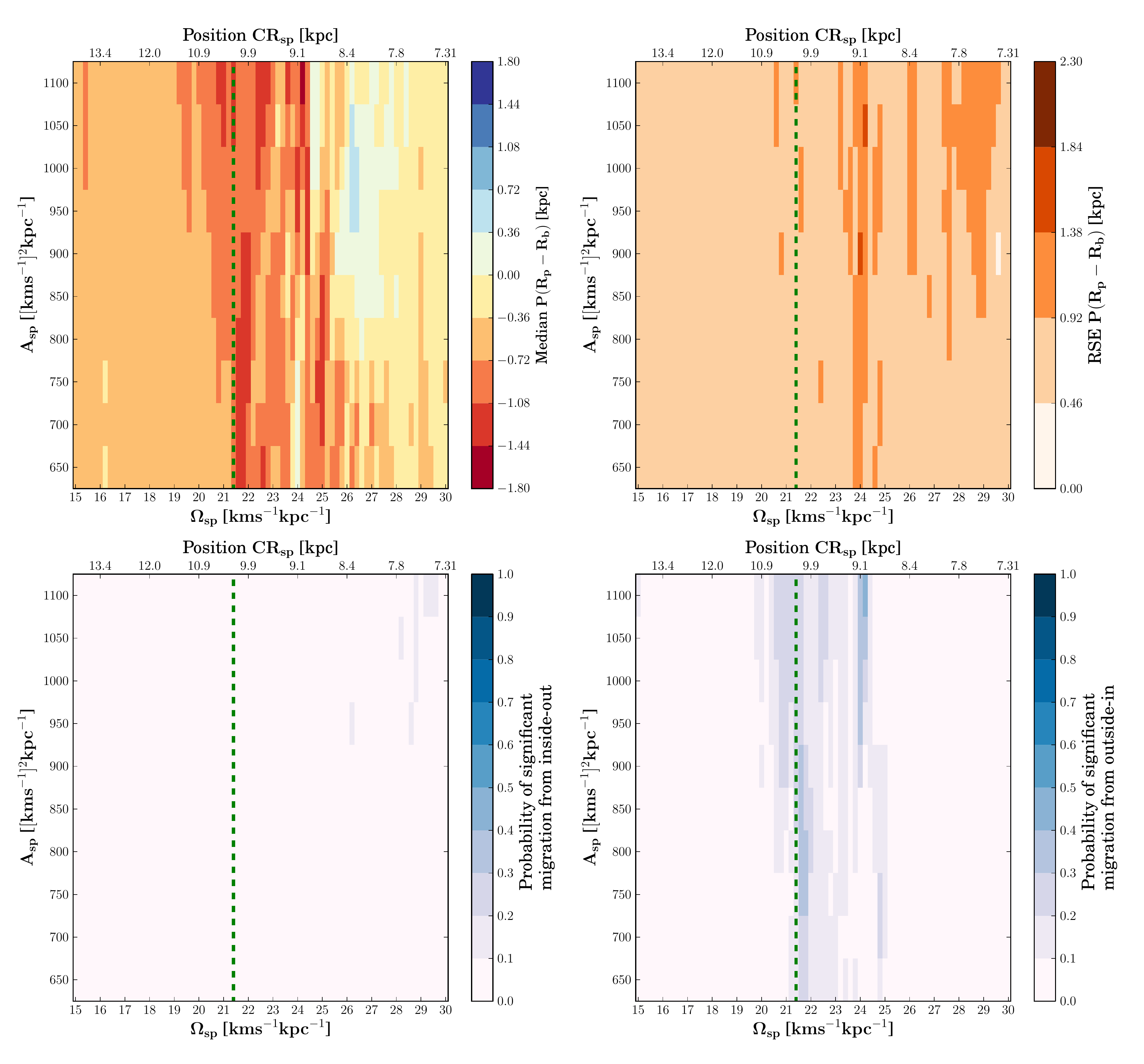}
  \caption{\textbf{Top:} Median and RSE of the distribution \pmigration\ as a function of
  the amplitude and pattern speed of a two-armed spiral structure. The location of the
  \crsp\ with respect to the Galactic centre is also shown. For this set of
  simulations, the position of the outer Lindblad resonance of the bar, \olrbar\ is fixed at
  $10.2$ kpc and it is shown as the vertical dotted green line. \textbf{Bottom:} \probio\ and
  \proboi also as a function of the amplitude and pattern speed of two spiral
  arms.\label{fig:ri_spiral_effect_m2}}
\end{figure*}

In this Section we study the effects of the spiral structure on the radial migration of the Sun and
thus keep fixed the mass and pattern speed of the bar. We chose the lowest limit for the bar mass
$M_\mathrm{bar}= 9.8\times10^9$ $M_\odot$. The pattern speed of the bar was set to be
$\Omega_\mathrm{bar}=40$ \patspeed. With this value, the resonances of the bar are located at
extreme regions in the Galactic disk, in particular \olrbar\ which is at $10.2$ kpc. In
Sect.\ \ref{sec:m2} and \ref{sec:m4}, we explore the effects of the amplitude, \crsp\ location and
number of spiral arms on the radial migration of the Sun.

\subsubsection{Effect of two spiral arms }\label{sec:m2}

\begin{figure*}
  \centering
  \includegraphics[width=15cm, height=20 cm]{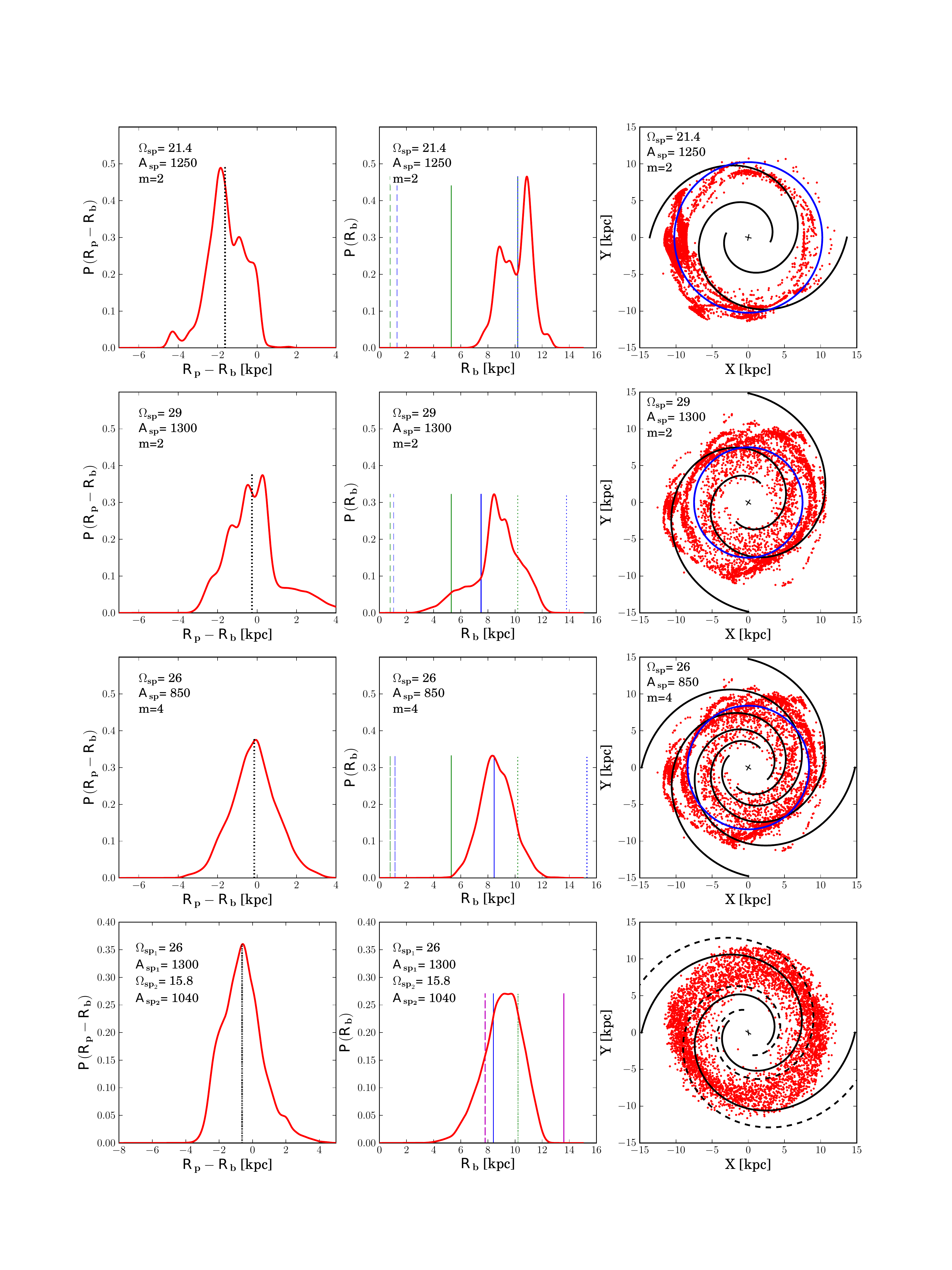}
  \caption{\textbf{Left:} Migration distribution \pmigration\ . \textbf{Middle:} distribution of possible Sun's birth radii $P(R_\mathrm{b})$. \textbf{Right:} Projection on the $xy$ plane of the possible birth radii of the Sun. A specific combination of bar and spiral arm parameters are used. In the first and second rows the Galactic potential has two spiral arms. In the third row, the Galactic potential has four spiral arms. In the bottom panel, we use a superposition of two spiral arms ($2+2$) with different pattern speeds. 
The pattern speed and mass of the bar are fixed to $\Omega_\mathrm{bar}= 40$ \patspeed\ and $M_\mathrm{bar}= 9.8\times10^9$ $M_{\odot}$, respectively.  The vertical dotted black line in the panels of the left is the
    median of the distribution \pmigration. The same line styles as in Fig.\ \ref{fig:ri_bar_effect}
    are used to indicate the resonances due to the bar and spiral arms in the panel of the middle. In the same panel at the bottom, the dashed and solid magenta lines correspond to the  \ilrsp\ and \crsp\ of the secondary spiral structure.
    The blue circle in the panels on the right in the first three rows represents the position of \crsp , which is located
    from top to bottom at $9.9$, $7.6$ and $8.4$ kpc respectively. The \crsp\ due to the multiple spiral patterns are not shown in the plot of the bottom. In the right panel we also show the configuration of the spiral arm potential $4.6$ Gyr ago. The dashed black line in the figure of the bottom represents the secondary spiral structure. \label{fig:xy_sp} }
\end{figure*}

In Fig.\ \ref{fig:ri_spiral_effect_m2} we show the characteristics of the migration distribution as
a function of the amplitude and pattern speed of two spiral arms. We varied the amplitude in steps
of $50$ \spiralampl\ and the pattern speed in steps of $0.2$ \patspeed. Note that for most of the
spiral arm parameters the median of \pmigration\ is negative, suggesting that the migration of the
Sun has been mainly from outer regions of the Galactic disk to $R_\odot$. If the \crsp\ is located
between $9.0$ and $10.6$~kpc with respect to the Galactic centre, the median of \pmigration\ remains between $-1.08$ and $-1.44$~kpc for most of the values of $A_\mathrm{sp}$.  The  median of \pmigration\ can reach values of up to $-1.80$~kpc if $A_\mathrm{sp}=1100$ \spiralampl\  and $\Omega_\mathrm{sp}= 24.2$ \patspeed\ (\crsp\ at $9$~kpc). For this latter case, there is  a probability between 40\% and 50\% that the Sun has migrated considerably from outer regions of the Galactic disk to its current position (cf.\ Fig.\ \ref{fig:ri_spiral_effect_m2}, bottom right panel).  

We also studied the radial migration of the Sun for amplitudes higher than $1100$ \spiralampl\,, up to $1300$~\spiralampl\ . We found that the migration of the Sun on average is from outer regions of the Galactic disk to $R_\odot$.  The Sun only migrates considerably when  $1200\leq A_\mathrm{sp}\leq 1300$ \spiralampl\ and $\Omega_\mathrm{sp} = [21.4,21.8]$ \patspeed\ . (\crsp\ $\sim 10.23$~kpc). According to the former results and given that  the \olrbar\ is located at $10.2$~kpc, the significant radial migration of the Sun occurs when  the distance between \crsp\ and \olrbar\ is in the range $[0 ,1]$ kpc. An illustration of the migration distribution \pmigration\ for these higher amplitudes is shown at the first and second rows of figure \ref{fig:xy_sp}. 

On the other hand, according to the bottom left panel of Fig.\  \ref{fig:ri_spiral_effect_m2}  we find that it is unlikely that the Sun has migrated from inner regions of the Galactic disk to $R_\odot$.

Other studies have also evaluated the effect of the spiral arms of the Milky Way on the motion of
the Sun. \cite{klacka} found that under the simultaneous effect of the central bar and spiral arms,
the Sun could experience considerable radial migration when it co-rotates with spiral arms that have
a strength $\epsilon=0.06$.  In our simulations this strength corresponds to  an amplitude $A_\mathrm{sp}= 1300$~\spiralampl\ .  According to our simulations the Sun experiences considerable radial migration when $A_\mathrm{sp}= 1300$~\spiralampl\ and $\Omega_\mathrm{sp}= [21.4,21.8]$ \patspeed\ ; therefore  significant radial migration is found when $\Omega_\mathrm{sp}=1.2\Omega_\odot$.

By comparing Fig.\ \ref{fig:ri_spiral_effect_m2} and \ref{fig:ri_bar_effect} we can see that a
2-armed spiral pattern tends to produce more radial migration on the Sun than the central bar of the
Milky way. \cite{sellwood}, and more recently \cite{minchev10}, found that the larger changes in
angular momentum of stars always occur near the co-rotation resonance, the effect of the outer/inner
Lindblad resonances being smaller.  Given that in our simulations the motion of the Sun is
influenced by the \crsp\ and by the \olrbar,  it is expected that the spiral arms produce a
stronger effect on the Sun's radial migration than the central bar of the Galaxy.

At the top panel of Fig.\ \ref{fig:xy_sp} we show the distributions \pmigration\  and
$p(R_\mathrm{b})$ for an example of a two-arm spiral arm potential that leads to considerable radial
migration of the Sun. In this case the distance between the \crsp\ and \olrbar\ is $0.03$ kpc. For this specific set of bar and spiral arm parameters the Sun could have migrated a distance
of  $1.8$ kpc from the outer regions of the Galactic disk to its current position. Its birth
radius would then be around 11 kpc, as also indicated by the distribution $p(R_\mathrm{b})$. The
projection of the Sun's birth locations in the $xy$ plane shows lots of structure, but again only
the birth radius can be constrained.

In the second row of Fig.\ \ref{fig:xy_sp} we show the distributions \pmigration\ and
$p(R_\mathrm{b})$  for a set of spiral arm parameters that produce high dispersion in the migration
distribution \pmigration\ . In this case the Sun does not migrate on average (Median
$\pmigration  \sim 0$). Additionally, as can be observed in the plot of the right,  there is a fraction of possible birth radii at the inner regions of the Galactic disk; however, the probability of significant migration from inside-out in this case  is only of  10\%.

\subsubsection{Effect of four spiral arms }\label{sec:m4}

\begin{figure*}
  \centering
  \includegraphics[width= 18 cm, height= 15cm]{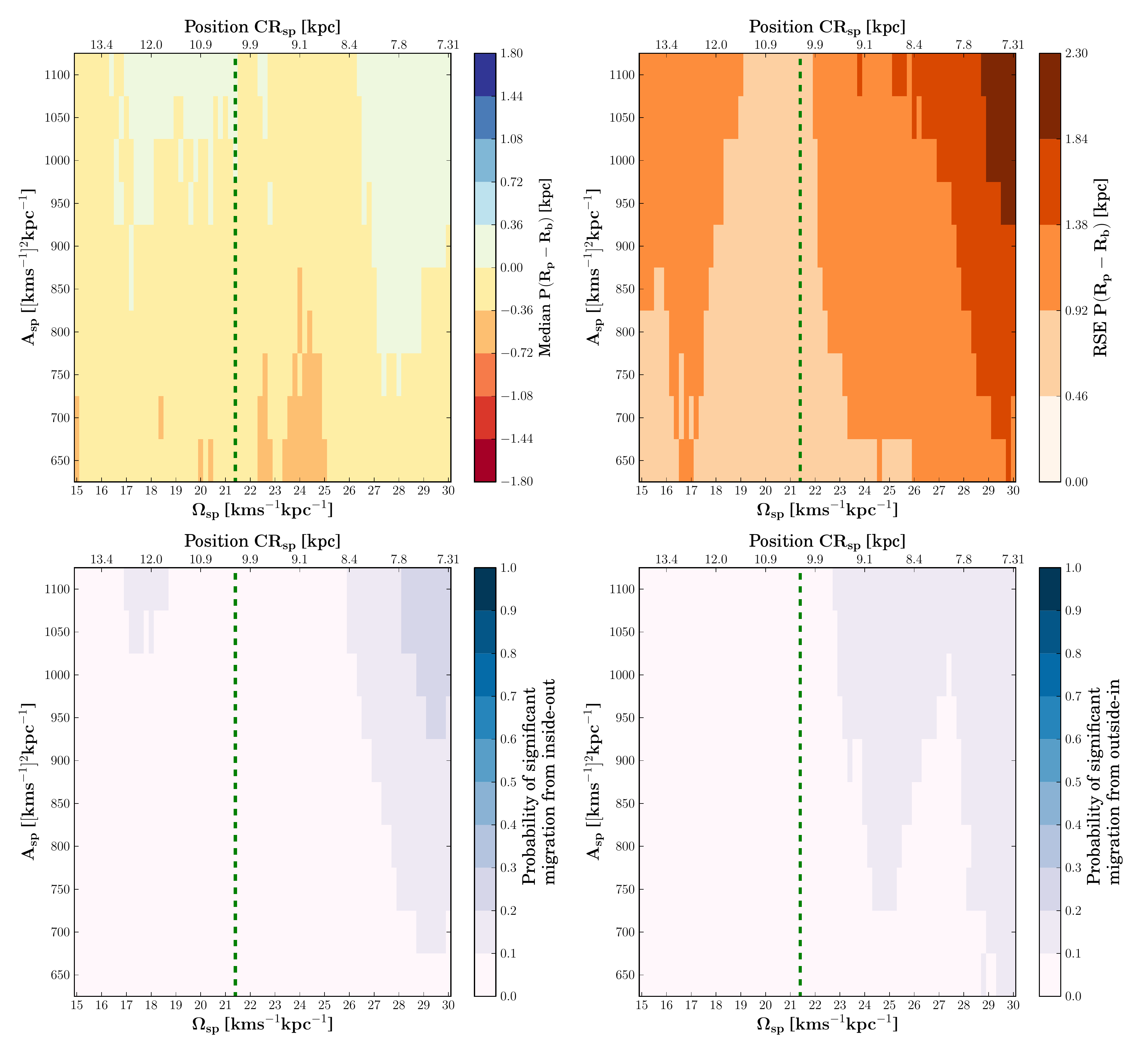}
  \caption{\textbf{Top:} Median and RSE of the distribution \pmigration\ as a function of the
  amplitude and pattern speed of a four-armed spiral structure. The location of the \crsp\
  with respect to the Galactic centre is also shown. For this set of simulations, the position of
  the \olrbar\ is fixed at $10.2$ kpc and it is shown as the vertical dotted green line.
  \textbf{Bottom:} \probio\ and \proboi\ also as a function of the amplitude and pattern speed of
  four spiral arms.\label{fig:ri_spiral_effect_m4}}
\end{figure*}

We also assess the radial migration of the Sun under the action of a Galactic potential composed of
four spiral arms. The results are shown in Fig.\ \ref{fig:ri_spiral_effect_m4}.  Note that when  $\Omega_\mathrm{sp}$ is between 19 and 22 \patspeed\, the radial migration experienced by the Sun is less than 1 kpc. Additionally, when \crsp\ is located between $7.3$
and $8.4$~kpc the median of \pmigration\ is between $-0.36$ and $0.36$ kpc (around zero). However
the large width of the distribution leads to probabilities of up to 30\% that the Sun has migrated
from inner regions of the Galactic disk to its current position.  The probability of significant migration in the other direction is up to 20\%.

The larger width of \pmigration\ may be due to the effect of higher order resonances (4:1
\ilrsp/\olrsp) on the motion of the Sun. The fact that the width of \pmigration\ is large for
specific four-armed Galactic potentials, means that the migration of the Sun is very sensitive to
its  birth phase-space coordinates. This effect can be also observed in the third row of
Fig.\ \ref{fig:xy_sp}, which shows \pmigration\ and $p(R_\mathrm{b})$ when the Galactic potential has
four spiral arms. In addition, the projection of the possible birth locations on the $xy$ plane shows
virtually no structure.

By comparing Figs.\ \ref{fig:ri_spiral_effect_m2} and
\ref{fig:ri_spiral_effect_m4}, we can see that unlike the case when the Galactic potential has two
spiral arms, the median of \pmigration\ when $m=4$ is not much affected by small separation
distances between the \crsp\ and \olrbar.

\subsubsection{Effects of multiple spiral patterns} \label{sec:lepine}

In addition to evaluating the motion of the Sun  in a pure 2-armed or 4-armed spiral structure, we use a superposition of two spiral arms ($2+2$) with different pattern speeds, such as discussed by \cite{lepine11}. We use the same values as  \cite{mishurov11}  to set the pitch angles of the multiple spiral patterns in the Milky Way.  The parameters of the main spiral structure used in the simulations are:  $ A_{\mathrm{sp}_1} = 650, 1300$~\spiralampl\ ;  $i_1 = -7^\circ$ and $\Omega_{\mathrm{sp}_1} = 26$~ \patspeed\ . This pattern speed  places the \crsp\ of the main spiral structure at  solar radius. The orientation of the main spiral pattern at the beginning of the simulations is $20^\circ$. 

The parameters used to model the secondary spiral structure are:  $A_{\mathrm{sp}_2}= 0.8A_{\mathrm{sp}_1}$;  $i_2= -14^\circ$ and $\Omega_{\mathrm{sp}_2}= 15.8$~\patspeed\  This  pattern speed places the \crsp\  of the secondary spiral structure at $13.6$~kpc and the 4:1~\ilrsp\ at $7.8$~kpc.  The orientation of the secondary spiral arms with respect to the main structure at the beginning of the simulations  is $-200^\circ$. In addition, we fixed the mass and pattern speed of the bar to $M_\mathrm{bar}= 9.8\times 10^9$ $M_\odot$ and $\Omega_\mathrm{bar}= 40$~\patspeed\ respectively. 

At the bottom panel of Fig. \ref{fig:xy_sp} we show the distributions  \pmigration\ and $P(R_\mathrm{b})$ when the Galactic potential has multiple spiral patterns. In this simulation the amplitude of the main spiral structure is  $A_{\mathrm{sp}_1}=1300$~\spiralampl\ . We used the tangential velocity of the Sun from \cite{bovy12}.  As can be seen,  the median of the distribution \pmigration\ is smaller than $1$~kpc, meaning that the migration of the Sun on average is not significant. The birth radius of the Sun is therefore at $8.5$~kpc, as can also be seen from the distribution  $P(R_\mathrm{b})$.  The projection of birth locations of the Sun  on the $xy$ plane  suggest that there is some fraction of possible birth radii located at internal regions of the Galactic disk; however, we found that the probability of considerable migration from outer or inner regions to $R_\odot$ is  between 8\% and 13\%. These probabilities are even smaller when $A_{\mathrm{sp}_1}=650$~\spiralampl\ .  We obtain the same results when assuming $V_\odot$ from \cite{schonrich}. 

In Sect.\ \ref{sec:m2} we have shown that the Sun might have experienced
considerable migration in the Galaxy if the \crsp\ is separated from the \olrbar\ by a distance smaller
than $1.1$ kpc. In the next Section we explore in more detail the effect of the bar-spiral arm
resonance overlap on the motion of the Sun.

%%%%

\subsection{Radial migration of the Sun in the presence of the bar-spiral arm resonance
overlap}\label{sec:overlap}

\begin{figure}
  \centering
  \includegraphics[width= 8.5 cm, height= 8 cm]{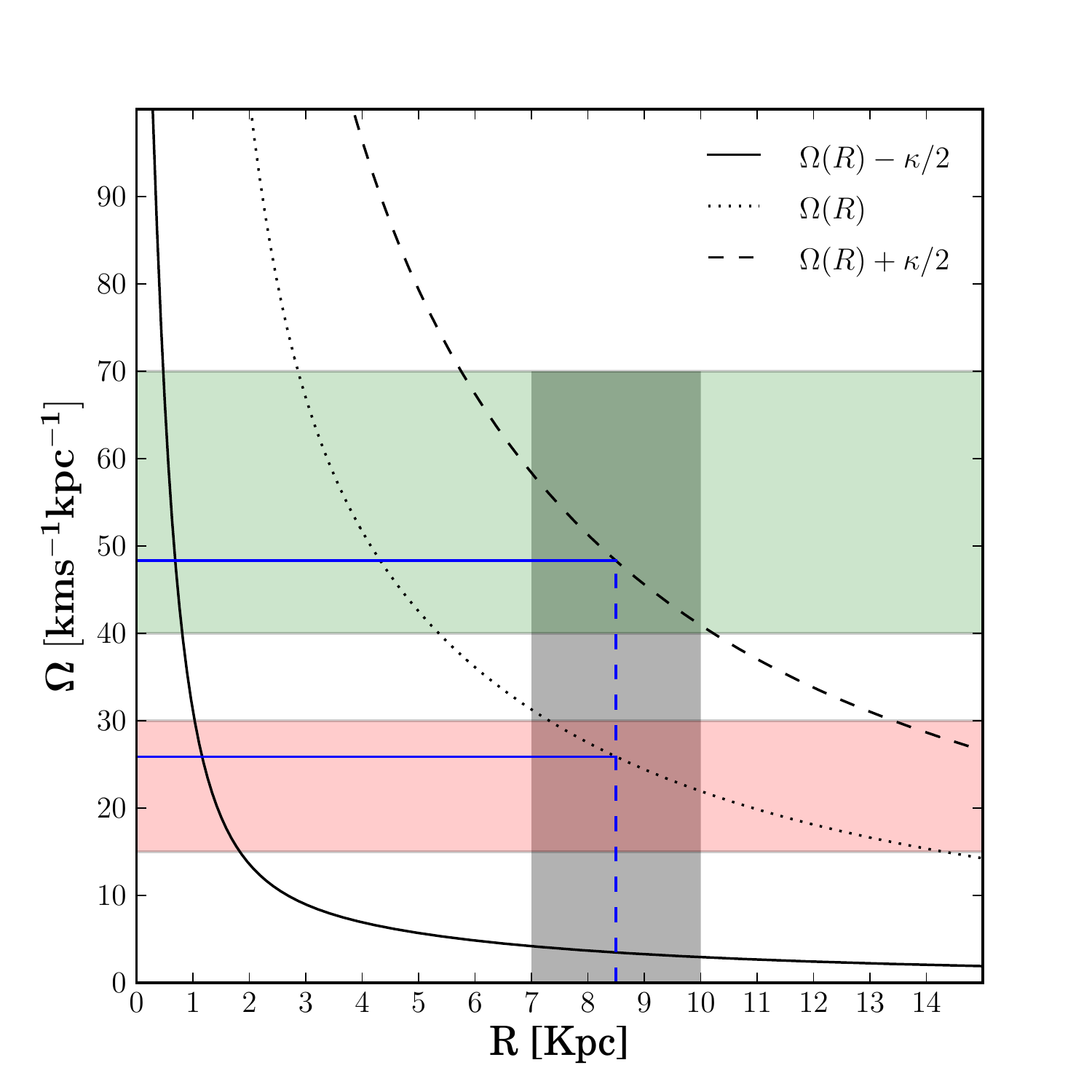}
  \caption{Resonances of second multiplicity (for $m=2$) in galactic disks.  The inner and outer
    Lindblad resonances (ILR, OLR) are along the solid and dashed black lines. They are given by:
    $\Omega(R) \pm \kappa/2$, where the minus (plus) sign corresponds to the ILR (OLR).  The
    corrotation resonance (CR) is along the dotted black line and  it is given by: $\mathrm{CR}=
    \Omega(R)$.  The shaded green region corresponds to the pattern speed of the bar within its
    uncertainty. The shaded red region corresponds to the pattern speed of spiral arms within its
    uncertainty.  Note that $\Omega_\mathrm{bar}$ and $\Omega_\mathrm{sp}$ only allow the
    overlapping between the Outer Lindblad resonance of the bar ($\mathrm{OLR_{bar}}$) with the
    corrotation of spiral arms ($\mathrm{CR_{sp}}$). We refer this resonance overlap as OLR/CR
    overlap. The gray shaded region is the location of the OLR/CR overlap  in the simulations. The
    blue lines show how we set  $\Omega_\mathrm{bar}$ and $\Omega_\mathrm{sp}$ to generate the
    OLR/CR overlap at some desired position.  \label{fig:overlap} }
\end{figure}

It has been demonstrated by \cite{minchev10} and \cite{minchev11} that the dynamical effects of
overlapping resonances from the bar and spiral arms provide an efficient mechanism for radial
migration in galaxies. Depending of the strength of the perturbations, radial mixing in Galactic
disks proceeds up to an order of magnitude faster than in the case of transient spiral arms. Given
that the solar neighbourhood is near to the  \olrbar\ and that the Sun is located approximately
at 1 kpc from \crsp\ \citep{acharova11}, it is of interest to study the radial migration
that the Sun might have experienced under the influence of the spiral-bar resonance overlap.

It is well known that galactic disks rotate differentially. However, the gravitational
non-axisymmetric perturbations such as the central bar and spiral arms, rotate as rigid bodies. In
consequence, stars at different radii will experience different forcing due to these
non-axisymmetric structures \citep{minchev10}. There are specific locations in the Galactic disk
where stars are in resonance with the perturbations. One is the corrotation resonance, where stars
move with the same pattern speed of the perturber,  and the Lindblad resonances, where the frequency
at which a star feels the the force due to the perturber coincides with its epicyclic frequency
$\kappa$.  Depending on the position of the star, inside or outside from the corrotation radius, it
can feel the Inner or Outer Lindblad resonances. 

In Fig.\ \ref{fig:overlap} we show the resonances of second multiplicity (for $m=2$) in a galactic
disk.  The green and red shaded regions correspond to the accepted values of the pattern speed of the bar and spiral arms of the Milky Way within the uncertainties.  As can be seen,  $\Omega_\mathrm{bar}$ and
$\mathrm{\Omega_{sp}}$  only allow certain combinations of resonance overlaps. For the case of two
spiral arms, only the overlap of the \olrbar\ and \crsp\ is possible \footnote{For $m=2$, we do not
  take into account second-order resonances, i.e. 4:1 $(\mathrm{ILR_{bar,sp}}, \mathrm{OLR_{bar,
  sp}})$}. Hereafter we refer to this resonance overlap as the OLR/CR overlap.

To explore the motion of the Sun in the presence the overlapping of resonances, we vary the pattern
speed of the bar and spiral arms such that the OLR/CR overlap is located at different positions in
the disk, between $7$ and $10.2$ kpc from the Galactic centre, as indicated by the vertical gray
shaded line in Fig.\ \ref{fig:overlap}. In our simulations, we varied the location of the OLR/CR
overlap every $0.1$ kpc. The amplitude of the spiral arms and the mass of the bar were also varied.

\begin{figure*}
  \centering
  \includegraphics[width= 18 cm, height= 10cm]{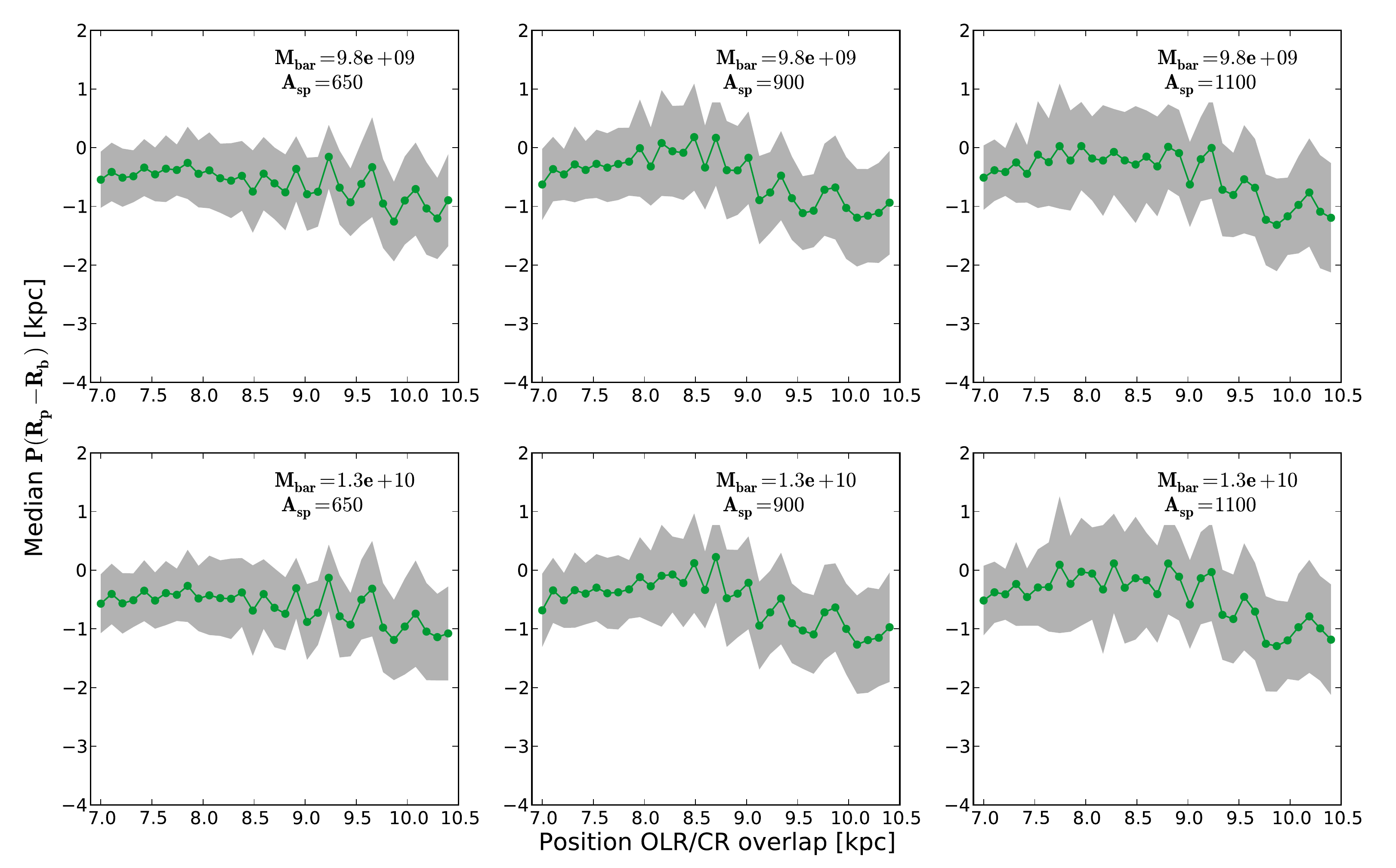}
  \caption{Median of the migration distribution \pmigration\ as a function of the position within
    the Galactic disk of the OLR/CR overlap. The shaded region corresponds to the RSE of the same
    distribution.  From left to right, the amplitude of the spiral arms, $A_\mathrm{sp}$ takes the
    values $650$, $900$ and $1100$ \spiralampl. From top to bottom, the mass of the bar,
    $M_\mathrm{bar}$ is $9.8\times10^9$ and $1.3\times10^{10}$ $M_\odot$.
    \label{fig:ri_resonance_overlap} }
\end{figure*}

\begin{figure*}
  \centering
  \includegraphics[width= 18 cm, height= 10cm]{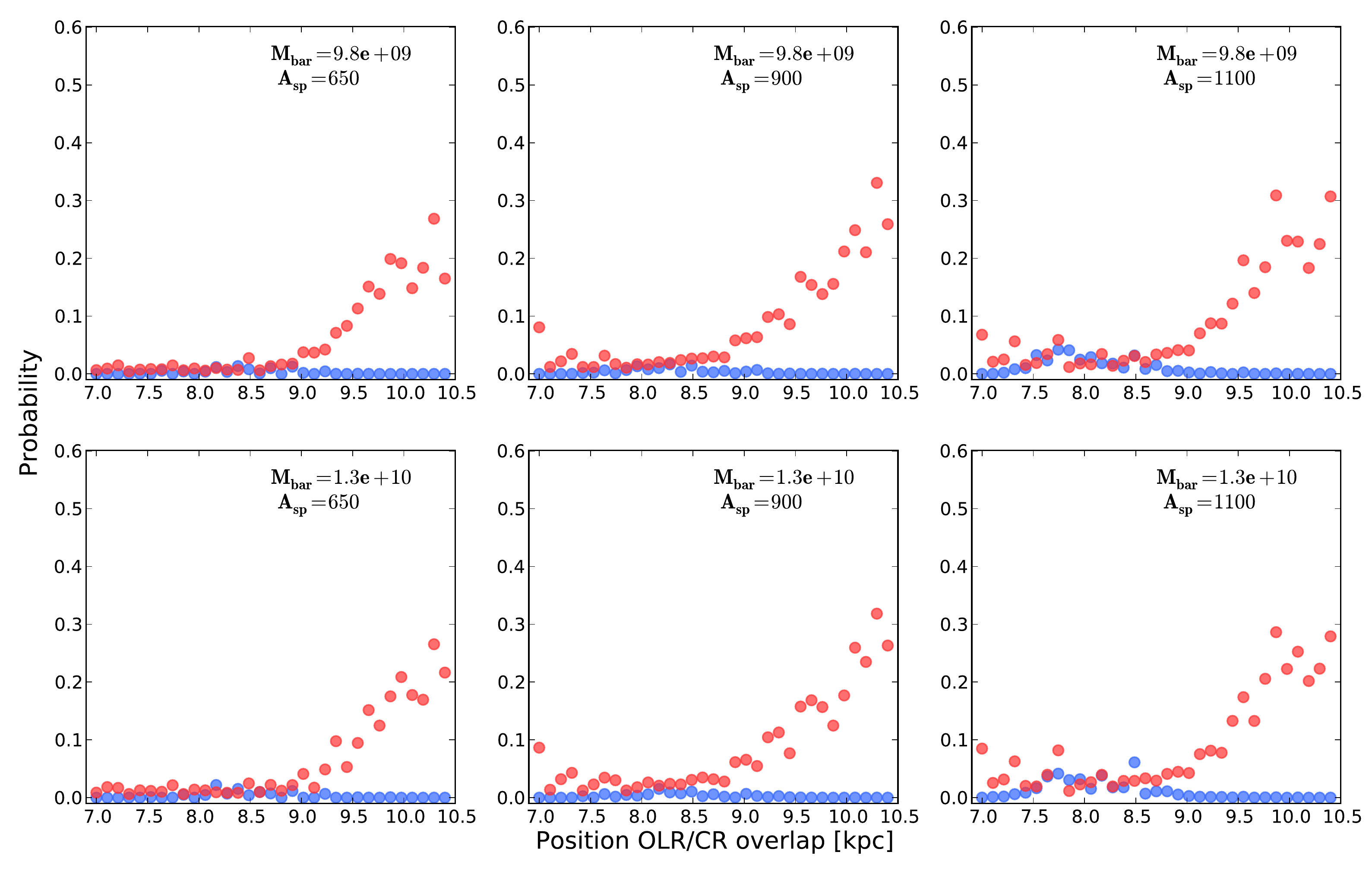}
  \caption{ Probability of considerable radial migration of the Sun as a function of the location of
    the OLR/CR overlap. The blue points represent the probability of significant migration of the
    Sun from inside-out \probio, while  the red points represent the significant migration from
    outside-in \proboi\ . The mass of the bar and amplitude of spiral arms are the same as in Fig.\
    \ref{fig:ri_resonance_overlap}. \label{fig:p_resonance_overlap} }
\end{figure*}

In Fig.\ \ref{fig:ri_resonance_overlap} we show the median of \pmigration\  as a function of the
position within the Galactic disk of the OLR/CR overlap. From left to right, the amplitude of spiral
arms increases; from top to bottom, the mass of the bar is $9.8\times10^9$ and $1.3\times10^{10}$
$M_\odot$.  Note that regardless of the amplitude of the spiral arms or the mass of the bar,  when
the OLR/CR overlap is located at distances smaller than $8.5$ kpc, the migration of the Sun is not
considerable. In fact, for these cases,  the probability that the Sun has migrated significantly in
either direction is smaller than $10$\% (see Fig.\ \ref{fig:p_resonance_overlap}). In contrast when
the OLR/CR overlap is located at distances larger than $8.5$ kpc, the median of the distribution
\pmigration\ is shifted towards negative values, while the probability for considerable migration
from the outer disk to $R_\odot$ goes up reaching  values up to $35$\%. The probability of significant migration from the inner disk to $R_\odot$ remains low at values of at most a few per cent.

\begin{figure}
  \centering
  \includegraphics[width= 5cm, height= 14cm]{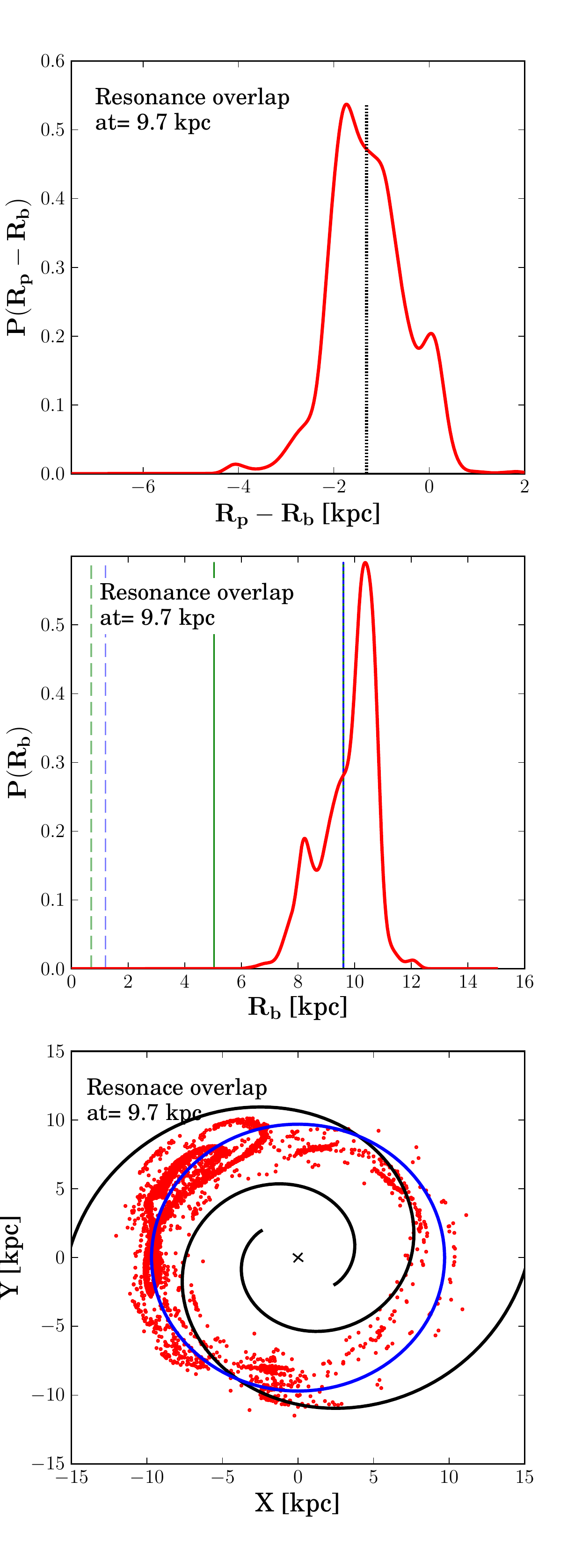}
  \caption{Example of the migration distribution for the case of the OLR/CR overlap located at 9.7 kpc
    from the Galactic centre. Here $A_\mathrm{sp}= 1100$ \spiralampl\  and
    $M_\mathrm{bar}=9.8\times10^{9}$ $M_\odot$. \textbf{Top:} The migration distribution
    \pmigration. The dotted black line indicates the median of the distribution. \textbf{Middle:}
    Distribution of the birth radius of the Sun $p(R_\mathrm{b})$. \textbf{Bottom:} Distribution of
    birth positions of the Sun projected on the $xy$ plane. The location of the OLR/CR overlap is
    indicated by the blue circle. The configuration of the spiral arm potential $4.6$ Gyr in the
    past is also shown.\label{fig:xy_resonance_overlap}}
\end{figure}

In Fig.\ \ref{fig:xy_resonance_overlap} we show the migration distribution for an example of a case
where the OLR/CR overlap has a strong effect, being located at $9.7$ kpc from the Galactic centre. For
this particular case, $M_\mathrm{bar}= 9.8\times10^{9}$ $M_\odot$ and $A_\mathrm{sp}=1100$
\spiralampl . The median of \pmigration\ is at $-1.3$ kpc and thus the radius where the Sun was born
is around 10 kpc. The latter can also be seen in in the distribution $p(R_\mathrm{b})$.  Note how
the distribution of birth positions in the $xy$ plane is clustered between the second and third quadrants. This is also seen for other cases, when the OLR/CR overlap is located between $8.5$ and $9.5$
kpc. However for different OLR/CR distances the clustering is toward other quadrants in the Galactic
plane. Hence, taking the uncertainties in the OLR/CR location into account again only the birth
radius of the Sun can be constrained.

\subsection{Radial migration of the Sun with higher values of its tangential velocity} \label{sec:v26}

\begin{figure}
  \centering
  \includegraphics[width= 8.5cm, height= 10.5cm]{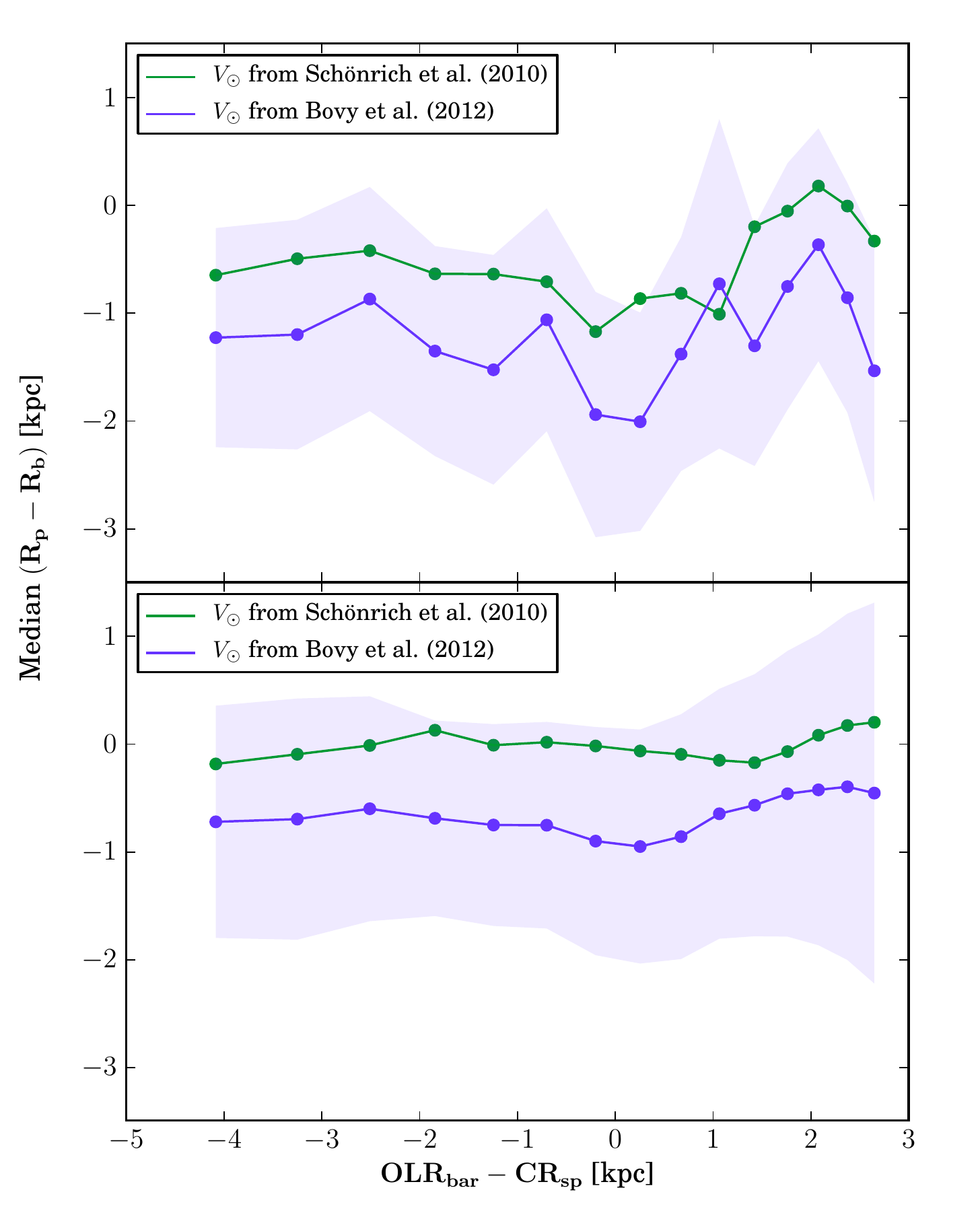}
 \caption{ Median of the distribution \pmigration\  as a function of the distance between the \olrbar\   and \crsp\ .  The green line is the resulting radial migration of the Sun when  we assume  a tangential velocity of $12.4 \pm 2.1$ kms$^{-1}$ \citep{schonrich}  in the orbit integration backwards in time.  The blue line is the radial migration of the Sun when we assume  a tangential velocity of $26\pm 3$~kms$^{-1}$ \citep{bovy12}.  The blue shaded region corresponds to the RSE of \pmigration\ for this latter case.  We used: \textbf{Top:}  two spiral arms. \textbf{bottom:} four spiral arms. \label{fig:other_VSun}   } 
 \end{figure}

In this section we explore the motion of the Sun backwards in time when assuming the rotational velocity suggested by \cite{bovy12}.  In Fig.\ \ref{fig:other_VSun} we show the median of the distribution \pmigration\ as a function of the distance between the \olrbar\ and \crsp\ . For this set of simulations we fixed the bar parameters to $M_\mathrm{bar}= 9.8\times10^9$~$M_\odot $ and $\Omega_\mathrm{bar}= 40$ \patspeed\  respectively. With this pattern speed, the \olrbar\ is located at $10.2$~kpc with respect to the Galactic centre.  Additionally, the amplitude of the spiral arms is  fixed to $A_\mathrm{sp}= 1050$ \spiralampl\ .  We varied the pattern speed of the spiral arms in steps of 1 \patspeed\  within the range listed in table \ref{tab:galparams}.  We used two and four spiral arms.  For comparison we have also plotted the median of the distribution \pmigration\ when the tangential velocity of the Sun  is taken from \cite{schonrich}.  As can be observed,  the migration of the Sun on average is approximately $1$~kpc higher when  $V_\odot$ is taken from \cite{bovy12}. In the latter case,  the median of the distribution \pmigration\  is negative  for both $m=2$ and $m=4$ meaning that the Sun has migrated from outer regions of the Galactic disk to $R_\odot$.  In addition, from the simulations shown at the top panel of Fig.\ \ref{fig:other_VSun}  we found that  when the \olrbar\ and \crsp\ are separated by $\pm0.2$~kpc, the Sun migrates on average a distance around $2$~kpc, placing the Sun's birth place at around $10.5$~kpc from the Galactic centre. For this specific case we found a probability between 55\% and 60\% that the Sun has migrated considerably from outer regions of the Galactic disk to its current position.  On the other hand, we found unlikely that the Sun has migrated from inner regions of the Galaxy to $R_\odot$. 

Contrary to the two-armed spiral structure, the migration of the Sun on average is not significant when $m=4$, even for small distances between the \olrbar\ and \crsp\ . (See bottom panel Fig.\ \ref{fig:other_VSun}). Note that the median of \pmigration\  is never greater than $-1.7$~kpc. However,  Given that the width of the distribution \pmigration\ is appreciable, specially when $\olrbar-\crsp\geq 2$~kpc, the probability of considerable migration from  inner or outer regions to $R_\odot$ can be of up to 10\% or 20\% respectively.

%%%%________________________________________
\section{Discussion}\label{sec:discussion}

It is well known that the metallicity of the interstellar medium (ISM) depends on time and Galactic
radius. Since younger stars formed at the same Galactocentric radius have higher metallicities, the
metallicity of the ISM is expected to increase with time. Additionally, it has been established that
the metallicity of the ISM decreases with increasing the Galactic radius due to more efficient star
formation and enrichment of the ISM in the central regions of galaxies \citep{daflon04, recio14}.

Past studies of the age-metallicity relation in the solar neighbourhood suggested that the Sun is
more metal rich by typically $0.2$ dex than most stars at its age and Galactocentric orbit
\citep{ed93, holmberg}.  Hence, from the relationship between metallicity and Galactocentric radius,
it is natural to deduce that the Sun might have migrated from the inner regions of the disk to its
current position in the Galaxy \citep{wielen96, minchev13}. However, if the observations are
restricted to stars within a distance of $40$ pc from the Sun it seems that its chemical composition
is not unusual after all. \cite{furman04} found a sample of 118 thin-disk stars with a mean age of
$4.5$ Gyr to have a mean metallicity of $-0.04$ dex. In addition \cite{valenti05}, found a mean
metallicity of $-0.01$ dex in a sample of F, G, and K stars that were observed in the context of
planet search programmes. More recently \cite{casagrande} found that the peak of the metallicity
distribution function of stars in the Geneva-Copenhagen survey \citep{nordstrom04}, is around the
solar value. As we mentioned in the introduction, if the Sun is indeed not more metal rich than the
stars of its same age and Galactocentric radius it is probable that the Sun has not experienced
considerable migration over its lifetime. 

\cite{minchev10} studied the effects of the bar-spiral arm resonance overlap in the solar
neighborhood. They found that  a large fraction of stars that were located initially at inner and
outer regions of the Galactic disk, ended up at a distance of $\sim 8$ kpc after  $3$ Gyr of
evolution. This explains  the observed lack of a  metallicity gradient with age in the solar
neighborhood \citep{haywood08}; however, the same simulations show that after $3$ Gyr of evolution,
the peak of the initial radial distribution of stars that end up at $8$ kpc is also around $8$ kpc,
meaning that most of the stars at solar radius, do not migrate. For their simulations,
\cite{minchev10} modelled the central bar and spiral arms of the Galaxy as non-transient
perturbations. 

In this study we find that large radial migration of the Sun is only feasible when the {\olrbar} is
separated from {\crsp} by a distance less than $\mathbf{1.1}$ kpc or when these two resonances overlap and
are located further than $8.5$ kpc from the Galactic centre.  In these cases we find that the
migration of the Sun is always from the outer regions of the Galaxy to $R_\odot$.  When the
$\mathrm{CR_{sp}}$ is located between $7.3$ and $8.4$ kpc and the number of spiral arms is four, the Sun migrates on average  little;
however, given that the width of the distribution \pmigration\ can be up to $2.3$ kpc, the radial
migration of the Sun highly depends on its birth phase-space coordinates. For this latter case, the
probability that the Sun has migrated considerably from inner regions of the disk to $R_\odot$ can be up to
30\% . Apart from the very
specific cases mentioned above, we found that in general the Sun might have not experienced
considerable radial migration from its birth place to its current position in the Galaxy.  In the
simulations we did not change the Galactic parameters (mass, scale length) of the axisymmetric
components of the Milky Way. Since this is a smooth potential, we do not expect great variations on
the solar motion due to the variation of these parameters. 

The model that we used for the Milky Way has two restrictions: it does not take into account
transient spiral structure and it assumes that the Galactic parameters have been fixed during the
last $4.6$ Gyr. Although there are several studies that suggest that the spiral structure in the
Galaxy is transient \citep{sellwood10, sellwood11}, the evolutionary history of the Milky Way is
quite uncertain, thus the Galactic model used is still valid.  The study of the radial migration of the Sun under the influence of transient spiral arms implies to extend the space of Galactic parameters even more. Hence, simulations taking into account transient spiral structure will be carried out in a future work.

Recently \cite{minchev13} made a more complex modeling of the Milky Way which involves
self-consistent N-body simulations in a cosmological context together with a detailed chemical
evolution model. They explored the evolution of a Galaxy for a time period of $11$ Gyr, which is
close to the age of the oldest disk stars in the Milky Way.  They found that as the bar grows and
the spiral structure start to form, the $\mathrm{CR_{bar}}$ and $\mathrm{OLR_{bar}}$ are shifted
outwards of the disk producing changes in the angular momentum of stars. These changes in angular
momentum can be doubled in the time interval from $4.4$ to $11.2$ Gyr.  At the end of the simulation
they found that stars of all ages end up at the solar neighborhood ($7\leq r\leq 9$ kpc).
Additionally, from the obtained metallicity distribution they conclude that the majority of stars
come from inner regions of the Galactic disk ($3\leq r \leq 7$ kpc), although a sizable fraction of
stars originating from outside the solar neighborhood is also observed. By assuming an error of
$\pm1$ dex in the metallicity, they found that the possible region where the Sun was formed is
between $4.4$ and $7.7$ kpc, with the highest probability to be around $5.6$ and $7$ kpc. These
results support the conclusions of \cite{wielen96}.

According to \cite{minchev13} the Sun probably has migrated a distance between $1.5$ and $2.9$ kpc
from the inner regions of the Galactic disk to $R_\odot$, which is different from what we obtained.
The discrepancy in the conclusions is due to the fact that the structure of the Milky Way and its
evolutionary history is quite uncertain. For instance,  \cite{minchev13} argued that their results are strongly
dependent on the migration efficiency in their simulations and also in the adopted chemical
evolution model. We obtained a broad  set of possible past Sun's orbits due to the large uncertainty
in the bar and spiral arm parameters. Hence, a large scale determination of the phase-space of stars
together with  better measurements of their chemical abundances  are needed to constrain the history
of the Milky Way and hence, their current properties.  With the {\em Gaia} mission \citep{gaia} we
can expect to obtain the parallaxes and proper motions of one billion of stars very accurately. The
{\em HERMES} \citep{hermes} and {\em APOGEE} \citep{apogee} surveys, will provide a complete
database of chemical abundances and radial velocities for stars across  all Galactic populations
(bulge, disk, and halo).    

With a more accurate determination of the Galactic parameters (masses, scale lengths, pattern
speeds), the motion of the Sun can be better constrained. 

\section{Summary and final remarks}\label{sec:summary}

We studied the radial migration of the Sun within the Milky Way by computing its past orbit in an analytical potential representing the Galaxy. We took into account the uncertainties in the
distance of the Sun from the centre of the Galaxy and its peculiar velocity components as well as
the uncertainties in the bar and spiral arm parameters.

At the start of the simulations the phase space coordinates of the Sun are initialized to 5000
different positions and velocities which were obtained from a normal distribution centred at
$(\vect{r}_\odot, \vect{v}_\odot)$, with standard deviations reflecting the uncertain present day
values of $\vect{r}_\odot$ and $\vect{v}_\odot$. After performing the backwards integration in time,
we obtained a distribution of `birth' phase-space coordinates.  We computed the migration
distribution function, \pmigration, to study the amount of radial migration experienced by the Sun
during the last $4.6$ Gyr. We obtain the following results: 

\begin{itemize}

  \item For the majority of the simulations the median of the distribution \pmigration\ is negative.
  This indicates that  the motion of the Sun has been on average from outer regions of the Galactic
  disk to $R_\odot$. 
  
\item The bar of the Milky Way does not produce considerable radial migration of the Sun. In
  contrast, the variation of amplitude and pattern speed of spiral arms produce migration on average
  of distances up to $-1.8$ kpc, if the number of spiral arms is two. Hence, the birth radius of the
  Sun would then be around $11$ kpc. In the case of a four-armed spiral potential, the Sun does not migrate on average; however, given that the width of the migration distribution \pmigration\ can be up to $2.3$~kpc, there is a probability of approximately 30\% that the Sun has migrated considerably from inner regions of the Galactic disk to $R_\odot$.  If the potential of the Galaxy has multiple non-transient spiral patterns, the Sun does not migrate on average.

\item Only very specific configurations of the Galactic potential lead to considerable migration of
  the Sun. One case is when the separation of the $\mathrm{OLR_{bar}}$ and $\mathrm{CR_{sp}}$ is
  less than or equal to $\mathbf{1}$ kpc.  Another case is when these two resonances overlap and are
  located further than $8.5$ kpc from the Galactic centre. For these cases there is a
  probability of up to $\mathbf{ 35}$\% or  $\mathbf{50}$\% that the Sun has experienced considerable radial migration from outer
  regions of the Galactic disk to $R_\odot$.

\item When the {\crsp} is located between $7.3$  and $8.4$~kpc and the Galactic potential has four spiral arms, the probability that the Sun has
  migrated considerably from inner regions of the Galactic disk to its current position can be up to
  30\%. For other combinations of bar and spiral arm parameters,
  $\probio \sim 0$. Hence, we found that in general it is unlikely that the Sun has migrated from inner regions of the Galaxy to $R_\odot$.

\item  Apart from the cases summarized above we find that in general the Sun might not have
  experienced appreciable migration from its birth place to its current position in the Galaxy.

\end{itemize}

In this study we consider the motion of the Sun in the plane. Simulations taking into account the vertical structure of the non-axisymmetric components of the Galactic potential  will be carried out in future works (e.g \cite{faure14, monari14} provide prescriptions for such potentials)

The study of the motion of the Sun during the last 4.6 Gyr has allowed us to determine its birth
radius. This is the first step to understand the evolution and consequent disruption of the Sun's
birth cluster in the Galaxy. In this respect, state of the art simulations are required to predict
more accurately the current phase-space of the solar siblings. In these simulations, internal
processes such as self gravity and stellar evolution have to be taken into account \citep{brown}.  A
detailed study of the evolution and disruption of the Sun's birth cluster by using realistic
simulations will appear in a forthcoming paper. 

According to the above results, the current distribution on the $xy$ plane of the solar siblings
will be different depending on the configuration of the Galactic potential. For the bar and spiral
arm parameters that produce a broad migration distribution \pmigration\ (cases where RSE $\geq 1.7$
kpc), we expect a high dispersion of solar siblings, spanning a large range of radii and azimuths on
the disk. For the Galactic parameters that do not generate a broad distribution \pmigration\ ( cases
where RSE $\leq 1$ kpc), we expect the Sun's siblings not to have a large radial dispersion.
Therefore, depending on their final distribution, it would be likely or unlikely to find solar
siblings in the near vicinity of the Sun.   

\cite{mishurov11} concluded that it is unlikely to find solar siblings within $100$ pc from the Sun,
since members of an open cluster are scattered over a large part of the Galactic disk when the
gravitational field associated to the spiral arms is taken into account. Consequently, a large scale
survey of phase-space is needed. Only the {\em Gaia} mission \citep{gaia} will provide data at the
precision needed to probe for siblings which are far away from the Sun \citep{brown}. The realistic
simulations mentioned above will have to be exploited to develop methods to look for solar siblings
among the billions of stars in the {\em Gaia} catalogue; however, together with the kinematics
provided by the simulations, a complete determination of chemical abundances of stars has to be done
to find the true solar siblings \citep{brown, ramirez12}.

The identification of the siblings of the Sun will enable to put better constraints on the initial
conditions of the Sun's birth cluster, instead of using only the current solar System properties
\citep{brown, adams}. With  well established initial conditions for the parental cluster of the Sun,
the formation, evolution and current features of the solar system could finally be disentangled. 

\section*{Acknowledgements}
We want to thank Merc\'e Romero G\'omez for providing us the model for the bar potential. Nathan de Vries and Inti Peluppesy for their valuable help that allowed us to make
the interface of the Galactic model and to develop the \rotbridge. We also want to thank the anonymous referee, Michiko
Fujii, Lucie Jilkov\'{a}, Kate Tolfree and Teresa Antoja for their suggestions and fruitful
discussions that improved this manuscript. This work was supported by the Nederlandse Onderzoekschool voor Astronomie (NOVA), the Netherlands Research Council NWO (grants \#639.073.803 [VICI],  \#614.061.608 [AMUSE] and \#612.071.305 [LGM]) and by the Gaia
Research for European Astronomy Training (GREAT-ITN) network Grant agreement no.: 264895

\bibliographystyle{mn2e}
\bibliography{references}

\appendix 
\section{A new approach for performing realistic simulations: Rotating Bridge}\label{app:bridge}
Spiral arms or bars rotate with some pattern speed, which means that the potential associated with
these Galactic components will depend on time in an inertial frame. Hence, in order to compute the
equations of motion of a set of stars in these Galactic components, it is convenient to chose a
frame which co-rotates with the bar or with the spiral arms so that the potential of these
perturbations will be time independent in the rotating frame. 

Let us consider a star in a frame that co-rotates with the central bar or with the spiral arms. The
Hamiltonian of this particle will be:
\begin{equation}\label{hamiltonian1}
  H= \frac{||\vect{p}||^2}{2m} + U_\mathrm{T}(\vect{r}) - \left( \Omega_\mathrm{p} \times \vect{r}
  \right)\cdot \vect{p}-\frac{1}{2}m ||\Omega_\mathrm{p} \times \vect{r}||^2\,,
\end{equation}

where \vect{r} and \vect{p} are the position and momentum vectors of the particle in the rotating
frame, $U_\mathrm{T}(\vect{r})= m(\phi_\mathrm{axi}(\vect{r}) + \phi_\mathrm{p}(\vect{r})$) is the
total potential energy due to the Galactic potential which is composed of an axisymmetric part
$\phi_\mathrm{axi}(\vect{r}) $ and a perturbation term $\phi_\mathrm{p}(\vect{r})$ that can be the
bar or spiral arms (where $m$ is the mass of the star). The last two terms of the Hamiltonian correspond
to a generalized potential energy due to the rotating frame, where $\Omega_\mathrm{p}$ is the
pattern speed of the bar or the spiral arms.

The above Hamiltonian can be written as:
\begin{equation}\label{hamiltonian}
  H= H_A + H_B\,,
\end{equation}
where
\begin{align*}
  H_A &= \frac{||\vect{p}||^2}{2m} \\
  H_B &= U_\mathrm{T}(\vect{r}) - m\left( \Omega_\mathrm{p} \times \vect{r}
  \right)\cdot\dot{\vect{r}} -\frac{1}{2}m ||\Omega_\mathrm{p} \times \vect{r}||^2\,.
\end{align*}

Note that $\vect{p}=m\vect{\dot{r}}$. A differential operator can be defined in terms of the Poisson
bracket: 
\begin{equation*}
  D_H= \{ \quad, H\}= \frac{\partial H}{\partial \vect{p}}\frac{\partial}{\partial \vect{r}} - \frac{\partial H}{\partial \vect{r}}\frac{\partial}{\partial \vect{p}}, 
\end{equation*}

so that the Hamilton's equations of motion can be written as:

\begin{subequations}
  \begin{align} 
    \vect{\dot{r}} &= \left( D_{H_A} + D_{H_B}\right) \vect{r} \label{r} \\
   \vect{ \dot{p}} &= \left( D_{H_A} + D_{H_B}\right) \vect{p} \label{p}.
  \end{align}
\end{subequations}

By solving eqs.\ \eqref{r} and \eqref{p} , we can express the time evolution of the position and
momentum of a particle:

\begin{subequations} 
  \begin{align} 
    \vect{r}(t+ \Delta t) &= e^{ \left( D_{H_A} + D_{H_B}\right) \Delta t} \vect{r}(t) \\
    \vect{p}(t+ \Delta t) &= e^{ \left( D_{H_A} + D_{H_B}\right) \Delta t} \vect{p}(t), 
  \end{align}
\end{subequations}

where $e^{ \left( D_{H_A} + D_{H_B}\right) \Delta t}$ is the time evolution operator that is defined as

\begin{equation}\label{evolution}
  e^{ \left( D_{H_A} + D_{H_B}\right)\Delta t} = \prod_{i=1}^{k} e^{ c_iH_A\Delta t} e^{ d_iH_B\Delta t} + O(\Delta t^{n+1}),
\end{equation}

where $n$ is an integer which corresponds to the order of the integrator. ($c_i$, $d_i$)
($i=1,2,\dots,k$) is a set of real numbers. The simplest case is when the integrator has second order.
This integrator, called Leapfrog, has the following coefficients: $n=2, c_1=0, d_1= \frac{1}{2},
c_2=1, d_2=\frac{1}{2}$. Those are the coefficients used in the interface of {\bridge} for \amuse.
In Sect.\ \ref{sect:high_order}, we will mention briefly the coefficients used in {\amuse} for high
order integrators. The operators in eq. \eqref{evolution} are applied to $r$ and $p$ in the order
($c_1,d_1,\dots, c_k,d_k$).

The kick (K) operator or $ e^{H_B\Delta t}$ produces the following set of equations:

\begin{subequations} 
  \begin{align}
    \vect{\dot{r}} &= 0 \label{rp} \\
   \vect{ \dot{p}} &= \vect{F} \label{pp},
  \end{align}
\end{subequations}

where $\vect{F}$ is the total force of the particle in the rotating frame, which is given by:

\begin{equation}
  \vect{F}= m\vect{a} - m\Omega_p\times \left(\Omega_p \times \vect{r} \right) - 2m\left(\Omega_p \times \vect{r} \right).
\end{equation}

Here $\vect{a}= -\nabla (\phi_\mathrm{axi}(\vect{r}) + \phi_\mathrm{p}(\vect{r}))   $. In case of having a central bar and spiral arms, one rotating frame has to
be chosen first to make the integration of the equations of motion. We chose a frame that co-rotates
with the bar. Every time step $\Delta t$, the force of the axisymmetric component plus bar is
computed; since spiral arms have a different pattern speed, the position of the star is calculated
in another frame that co-rotates with the spirals; there, the force due to this perturbation is
computed. Finally, the position goes back to the co-rotating frame with the bar to calculate the
total force.

Given that the momentum can also be written as $\vect{p}= m\vect{v}$, Eqs. \eqref{rp} and \eqref{pp} can be written as

\begin{subequations} 
  \begin{align} 
    \dot{x} &= \dot{y} = \dot{z} = 0 \\
    \dot{v}_{x} &= a_x + \Omega_p^2x + 2\Omega v_{y} \\
    \dot{v}_{y} &= a_y + \Omega_p^2y - 2\Omega v_{x} \\
    \dot{v}_{z} &= a_z
  \end{align}
\end{subequations}

The solution to this system of equations is
\begin{subequations} 
  \begin{align} 
    v_{x} (t+\Delta t) &= \left[ v_{x{}}(t) -\left(\frac{a_y + \Omega^2y}{2\Omega} \right) \right]\cos{(2\Omega\Delta t)} \nonumber \label{v1} \\
    & + \left[ v_{y{}}(t) +\left(\frac{a_x + \Omega^2x}{2\Omega} \right) \right]\sin{(2\Omega\Delta t)} \nonumber \\
    & + \frac{a_y+ \Omega^2y}{2\Omega} \\
    v_{y} (t+\Delta t) &= \left[ v_{x{}}(t) -\left(\frac{a_y + \Omega^2y}{2\Omega} \right) \right]\sin{(2\Omega\Delta t)} \nonumber \\
    & + \left[ v_{y{}}(t) +\left(\frac{a_x + \Omega^2x}{2\Omega} \right) \right]\cos{(2\Omega\Delta t)} \nonumber \\
    & - \frac{a_x+ \Omega^2x}{2\Omega}\\
    v_{z} (t+\Delta t) &= v_{z} (t) + a_z\Delta t \label{v3}
  \end{align}
\end{subequations}

On the other hand, the drift (D) operator or $ e^{H_A\Delta t}$, produces this set of equations:
 
\begin{subequations} 
  \begin{align} 
    \dot{x} &= p_{x} \\
    \dot{y} &= p_{y} \\
    \dot{z} &= p_{z} \\
    \dot{p}_{x} &= \dot{p}_{y}= \dot{p}_{z}= 0, 
  \end{align}
\end{subequations}

which give the solution:

\begin{subequations} 
  \begin{align}
    x(t+\Delta t) &= x(t) + v_{x}(t+ \Delta t)\Delta t \\
    y(t+\Delta t) &= y(t) + v_{y}(t+ \Delta t)\Delta t \\
    z(t+\Delta t) &= z(t) + v_{z}(t+ \Delta t)\Delta t. 
  \end{align}
\end{subequations} 

In the more general case of having a star cluster with self gravitating particles, its hamiltonian will be

\begin{align}
  H = & \sum_{i}^n\frac{||\vect{p}_i||^2}{2m_i} -\sum_{i<j}^n \frac{Gm_im_j}{||\vect{r}_{ij}||} + \sum_i m_i (\phi_\mathrm{axi}(\vect{r}_i) + \phi_\mathrm{p}(\vect{r}_i))
   \nonumber \\
  &- \sum_i \left( \Omega_p \times \vect{r}_{i} \right)\cdot \vect{p}_{i} -\frac{1}{2}\sum_i m_i ||\Omega_p \times \vect{r}_{i} ||^2,
\end{align}

which can be separated as equation \eqref{hamiltonian1} with the terms

\begin{align*}
  H_A &= \sum_{i}^n\frac{||\vect{p}_i||^2}{2m_i} -\sum_{i<j}^n \frac{Gm_im_j}{||\vect{r}_{ij}||} \\
  H_B &= \sum_i m_i (\phi_\mathrm{axi}(\vect{r}_i) + \phi_\mathrm{p}(\vect{r}_i)) - \sum_i \left( \Omega_p \times \vect{r}_{i} \right)\cdot \vect{p}_{i}
   \nonumber \\
  & -\frac{1}{2}\sum_i m_i ||\Omega_p \times \vect{r}_{i} ||^2. 
\end{align*}

For this system, the kick operator gives the same set of velocities as in Eqs. \eqref{v1}-
\eqref{v3}; nevertheless, when the drift operator is applied, additionally to the position, the
velocity of the particles has to be updated again by taking into account their gravitational
interaction, as is explained in Sect.\ 2 of \cite{bridge}. 

\begin{figure*}
  \centering
  \includegraphics[width= 18cm, height= 6cm]{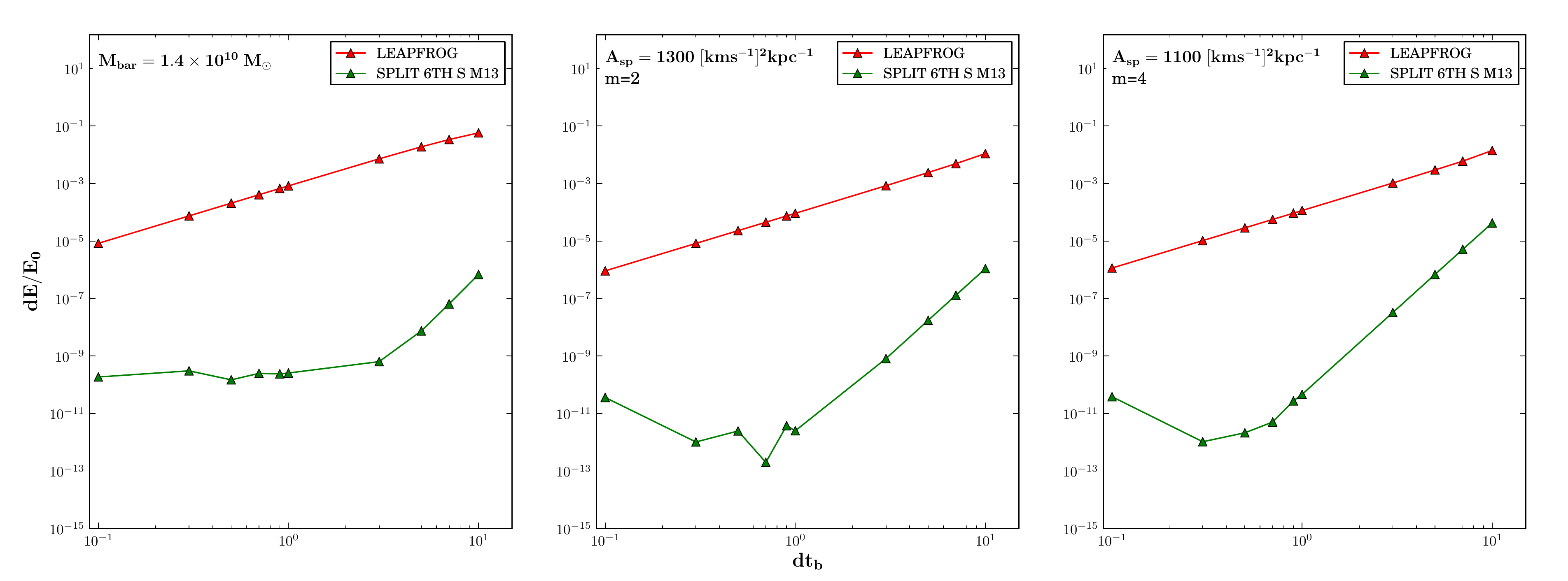}
  \caption{ Fractional energy error as a function of the {\rotbridge} time step when the Galactic potential is composed by: \textbf{Left:} Axisymmetric part + central bar. \textbf{Middle:} Axisymmetric part + two spiral arms. \textbf{Right:} Axisymmetric part + four spiral arms. The fractional energy error was computed by using a star with the following galactocentric initial position and velocity: $\vect{r}=(-8.5, 0, 0)$~kpc; $\vect{v}=(11.1, 12.24+V_\mathrm{LSR}, 0)$~km~s$^{-1}$. $V_\mathrm{LSR}$ is the velocity of the local standard of rest. \label{fig:energy_error}}
\end{figure*}

\subsection{High order integrators}\label{sect:high_order}

In AMUSE several high order integrators have been implemented from 4th until 10th order. In the
case of 4th order integrators, they can have 4,5 or 6 stages (named M4, M5 or M6); that is, the
number of times the force is computed when is applied Eq. \eqref{evolution}. The coefficients used
in those integrators are the ones found by \cite{mclachan} and \cite{mclachan1}. The 6th order
integrators implemented in AMUSE are of 11 and 13 stages (M11 and M13), with the coefficients found
by \cite{sofroniou}.  In the simulations performed here, we used the {\rotbridge} with a 6th order integrator. 

 In order to assess the accuracy of the code, we computed the energy error as a function of the {\rotbridge} time step ($dt_\mathrm{b}$) for a solar orbit under different bar and spiral arm parameters. The results are shown in Fig.\ \ref{fig:energy_error}.  Note that for a fixed $dt_\mathrm{b}$ the 6th order Leapfrog  can be six orders of magnitude more accurate than the second order Leapfrog. We found that such accuracy is independent from the bar and spiral arm parameters, as also can be seen from Fig.\  \ref{fig:energy_error}.

We chose a $dt_\mathrm{b}$ of 0.5 Myr for the simulations, which corresponds to a energy error of the order of $\mathbf{10^{-10}}$ when the Galactic potential has only a central bar or
spiral arms.

\label{lastpage}

\bsp

\end{document}